\documentclass[english,twoside]{article}
\usepackage[toc,page]{appendix}
\usepackage[utf8x]{inputenc} 
\usepackage{newunicodechar}
\usepackage{url} 
\usepackage{blindtext} 
\usepackage{titling}
\usepackage{pgfplots}
\usepackage{wrapfig}
\usepackage{graphicx}
\usepackage{pstricks}
\usepackage{float}
\pgfplotsset{width=7cm,compat=1.9}
\usepackage[many]{tcolorbox}
\usepackage{pst-all}
\usepackage{pstricks-add}

\usepackage[english]{babel}
\usepackage{amsthm}
\newtheorem*{theorem}{Theorem}

\newtheorem*{definition}{Definition}

\usepackage{tocbibind}
\usepackage[ruled,dotocloa]{algorithm2e}

\newtcolorbox[auto counter]{example}[1][]{
	top=.3cm,bottom=.3cm,
	enlarge top by=.4cm,
	fontupper=\sffamily\bfseries,
	colback=white, 
	colframe=black, 
	toptitle=1mm,
	bottomtitle=1mm,
	arc=0pt, 
	fonttitle=\bfseries,
	title={Example \thetcbcounter},
	label=#1
}

\newcommand\TBox[2][]{
	\tikz\node[draw,align=left,inner sep=2mm,#1] {#2};
}

\usepackage[english]{babel} 
\usepackage{varioref}
\addto\extrasenglish{
}

\usepackage[sc]{mathpazo} 
\usepackage[T1]{fontenc} 
\usepackage{microtype} 
\usepackage{alltt}

\usepackage[hmarginratio=1:1,top=32mm,columnsep=20pt]{geometry} 
\usepackage[hang, small,labelfont=bf,up,textfont=it,up]{caption} 
\usepackage{booktabs} 

\usepackage{lettrine} 

\usepackage{enumitem} 
\setlist[itemize]{noitemsep} 
\setlist[enumerate]{noitemsep} 

\usepackage{paralist}

\usepackage{abstract} 

\usepackage{titlesec} 
\renewcommand\thesection{\Roman{section}} 
\renewcommand\thesubsection{\roman{subsection}} 
\titleformat{\section}[block]{\large\scshape\centering}{\thesection.}{1em}{} 
\titleformat{\subsection}[block]{\large}{\thesubsection.}{1em}{} 

\usepackage{fancyhdr} 
\usepackage{titling} 

\usepackage{hyperref}
\usepackage{amssymb}
\usepackage{alltt}
\usepackage{lmodern}
\usepackage{tocloft}
\usepackage{etoolbox}

\renewcommand\thesection{\arabic{section}}
\renewcommand\thesubsection{\arabic{section}.\arabic{subsection}}

\title{Witnet: A Decentralized Oracle Network Protocol} 
\author{
	\textsc{Adán Sánchez de Pedro}\\[-5pt]
	\textsc{\small Stampery, CTO}\\[-5pt]
	\normalsize \href{mailto:adan@stampery.com?subject=Witnet Whitepaper}{adan@stampery.com} 
	\and
	\textsc{Daniele Levi}\\[-5pt]
	\textsc{\small Stampery, CEO}\\[-5pt]
	\normalsize \href{mailto:daniele@stampery.com?subject=Witnet Whitepaper}{daniele@stampery.com} 
	\and
	\textsc{Luis Iván Cuende}\\[-5pt]
	\textsc{\small Aragon, Project Lead}\\[-5pt]
	\normalsize \href{mailto:luis@aragon.one?subject=Witnet Whitepaper}{luis@aragon.one} 
}
\date{Version 0.1\\November 27, 2017}

\begin{document}
	\raggedbottom
	
	\pretitle{\begin{center}\begin{bfseries}\Large}
			\posttitle{\\ \large{[ Working Draft // Early Request for Comments ]}\footnote{\noindent\textbf{Note}: Witnet is a work in progress. Active research is under way, and new versions of this paper will appear at https://witnet.io. For comments and suggestions, contact us at \href{mailto:research@witnet.io?subject=Witnet Whitepaper}{research@witnet.io.}}\end{bfseries}\end{center}}
	\maketitle
	
	\renewcommand{\abstractname}{Abstract\vspace{-1em}}
	\begin{abstract}
		\noindent Witnet is a decentralized oracle network (\textsf{DON}) that connects smart contracts to the outer world. Generally speaking, it allows any piece of software to retrieve the contents published at any web address at a certain point in time, with complete and verifiable proof of its integrity and without blindly trusting any third party.
		
		\noindent Witnet runs on a blockchain with a native protocol token (called \textsf{Wit}), which miners---called \textit{witnesses}---earn by retrieving, attesting and delivering web contents for clients. On the other hand, clients spend \textsf{Wit} to pay witnesses for their \textsf{Retrieve-Attest-Deliver} (\textsf{RAD}) work. Witnesses also compete to mine blocks with considerable rewards, but Witnet mining power is proportional to their previous performance in terms of honesty and trustworthiness---this is, their reputation as witnesses. This creates a powerful incentive for witnesses to do their work honestly, protect their reputation and not to deceive the network.
		
		\noindent The Witnet protocol is designed to assign the \textsf{RAD} tasks to witnesses in a way that mitigates most attack vectors to the greatest extent. At the same time, it includes a novel 'sharding' feature that (1) guarantees the efficiency and scalability of the network, (2) keeps the price of \textsf{RAD} tasks within reasonable bounds and (3) gives clients the freedom to adjust certainty and price by letting them choose how many witnesses will work on their \textsf{RAD} tasks.
		
		\noindent When coupled with a Decentralized Storage Network (\textsf{DSN}), Witnet also gives us the possibility to build the Digital Knowledge Ark: a decentralized, immutable, censorship-resistant and eternal archive of humanity's most relevant digital data. A truth vault aimed to ensure that knowledge will remain democratic and verifiable forever and to prevent history from being written by the victors.
	\end{abstract}
	
	\newpage
	\tableofcontents
	
	\newpage
	\listoffigures
	
	\newpage
	\section{Introduction}
	\label{sec:introduction}
	
	\subsection{Motivation}
	\label{subsec:motivation}
	
	\lettrine{O}{ver} the last years, Bitcoin, Ethereum and other blockchain networks have proven the undeniable utility of decentralized transaction ledgers. These public ledgers process sophisticated smart contract applications that give digital money the capability to be programmed and incorporate logic. Those smart contracts, in their various forms, have the potential to allow for a more decentralized economy where individuals and companies worldwide can transact freely and safely without the need of intermediaries or trusted third parties.
	
	However, current smart contract solutions are self-contained in their supporting blockchains and have very little to no capability to interact with other blockchains, the Internet and the rest of the world. They were built deliberately detached from the outer world for a good reason: they need to be deterministic\footnote{In computer science, a deterministic algorithm is the one which, given a particular input, will always produce the same output, with the underlying machine always passing through the same sequence of states.}, while events in the real world are highly undeterministic if not completely random. 
	
	The most widely known way to feed outer information into smart contracts is using an oracle. An oracle is a trusted entity which signs claims about the state of the world. As signatures can be verified deterministically, oracles allow smart contracts to react to events happening in the outside world. But given that this approach puts trust in a single attesting third party (the oracle), it can not be considered trustless\footnote{In this work, like in most blockchain literature so far, the term \textit{trustless} has a very different meaning than the one in the dictionary. Instead of \textit{"not worthy of trust"}, it means \textit{"that operates without relying on blind trust in third parties"}.} nor tamper-resistant and therefore it leaves space to contestation and repudiation. It is not that the honesty and transparency of anyone is called into doubt. The problem here is that such a single source of truth also represents a single point of failure that introduces the chance for an external malicious actor to rewrite or delete facts by breaking into a single system or network.
	
	There is no way for smart contracts to deliver in their economic decentralization promises until we build some kind of decentralized oracle that does not rely on blind trust but on some digital equivalent of \textit{the wisdom of the crowd}.
	
	\subsection{Solution Overview}
	\label{subsec:overview}
	
	\lettrine{U}{nlike} the oracles that have traditionally been described, the one we are proposing here is not based on trust. Its claims can be considered to be true not because of the authority of the proposed network as a whole or that of any of its members in particular; but because the claims themselves are built by comparing and combining a number of likely conflicting claims coming from a plurality of anonymous players. Those players (the miners) have divergent interests but are strongly incentivized to tell the truth, which guarantees that they will not collude to deceive the system or feed any false claim into it.
	
	The proposed network can also be considered to work in a similar way to a decentralized prediction market\footnote{Prediction markets are exchange-traded markets created for the purpose of trading the outcome of events. They are based on the idea that market prices can indicate what the crowd thinks the probability of the event is. Notable examples of decentralized prediction markets are Augur\cite{augur}, Gnosis\cite{gnosis} and Delphi\cite{delphi}.} in the sense that the outcome of the attestations comes from the tally of the claims that a plurality of volunteer peers who are unknown to each other (the miners) have "voted for".
	
	Of course, the so-called miners are not actual human beings sitting in front of a computer, fulfilling assignments coming from an Internet overlord that commands them to use their web browser to navigate to a certain website and take a snapshot or copy some text that they must report. Instead, the miners are computers running a software that automatically receive and execute a series of tasks without the owner of the computer having to actively do anything else than install it and configure how much of the available CPU and bandwidth will be devoted to perform the tasks.
	
	Summarizing:
	
	\begin{itemize}
		\item The Witnet protocol is a Decentralized Oracle Network (\textsf{DON}) built on a blockchain with a native token. Clients spend tokens to get web contents retrieved, attested and delivered back to them; while a special kind of participants called \textit{witnesses} earn tokens for fulfilling such work.
		\item Witnesses are assigned tasks based on their previous performance, which is measured by reputation points. The more reputation a witness has, the more likely it will be assigned a task and the more its claims (solutions to assigned tasks) will be taken into account.
		\item Each witness earns reputation points when its claims match the claims brought by a majority of its peers. On the contrary, reputation points are deducted when its claims contradict the majority.
		\item Finally, witnesses can also participate in the creation of new blocks for the underlying blockchain. The likelihood that a witness will become a miner for the next block is proportional to its current reputation.
	\end{itemize}

	\subsection{Key Components}
	\label{subsec:components}
	
	The Witnet protocol builds upon a series of novel components:
	
	\begin{enumerate}
		\item \textbf{Decentralized Oracle Network} (\textsf{DON}): We provide an abstraction for network of independent providers to offer \textsf{Retrieve-Attest-Deliver} services. Later, we present the Witnet protocol as an incentivized, auditable and verifiable \textsf{DON} construction.
		\item \textbf{Reputation-Based Mining Protocol}: We show how to construct a useful Proof-of-Work that can be used in consensus protocols. Miners do not need to spend wasteful computation to mine blocks, but instead must fulfill task assignments.
		\item \textbf{Reputation-Based Task Assignment Protocol}: We put forward an algorithm that is analogous to the Reputation-Based Mining Protocol and lets the network assign tasks to miners in a decentralized, fair, uniform, unpredictable yet deterministic way.
		\item \textbf{Bridge nodes}: We describe a special kind of network participant that focus in the 'Deliver' part of the Retrieve-Attest-Deliver (\textsf{RAD}) tasks by allowing Witnet to interact with other blockchains.
	\end{enumerate}

	\newpage
	\section{Definition of a Decentralized Oracle Network}
	\label{sec:don}
	
	We outline a Decentralized Oracle Network (\textsf{DON}) as:
	
	\begin{itemize}
		\item a computer network made up of nodes (computers running a specific software),
		\item which communicate and operate as peers in compliance with an agreed protocol,
		\item to acquire knowledge of information that is external to the network,
		\item verify and agree on the veracity of the acquired information,
		\item and supply such verified information to other applications or networks that may need it.
	\end{itemize}
	
	In other words, a Decentralized Oracle Network is a \textit{peer-to-peer} (P2P) network capable of processing \textsf{Retrieve-Attest-Deliver} requests.
	
	\subsection{\textsf{Retrieve-Attest-Deliver} (\textsf{RAD}) Requests and Tasks}
	\label{subsec:rad}
	
	\textsf{Retrieve-Attest-Deliver} (\textsf{RAD}) requests are the primary and most important element in a \textsf{DON}. As their name suggests, they contain specific information on how a piece of information must be retrieved, attested and delivered.
	
	Going back to the outline of \textsf{DON}, the three elements of a \textsf{RAD} request can be easily defined:
	
	\begin{itemize}
		\item \textsf{Retrieve}: to acquire knowledge of information that is external to the network.
		\item \textsf{Attest}: to verify and agree on the veracity of the retrieved information.
		\item \textsf{Deliver}: to supply such attested information to the creator of the \textsf{RAD} request.
	\end{itemize}
	
	When a \textsf{DON} participant sends a \textsf{RAD} request to the network, the rest of participants must process the request by (1) retrieving the information, (2) check that they all have verbatim copies of the information, and (3) make it available for the the requester.
	
	When a \textsf{RAD} request is assigned to a \textsf{DON} participant, it is called a \textsf{RAD} task.
	
	In order to incentivize people to run nodes in a \textsf{DON}, the network can implement some form of native token that the participant sending a \textsf{RAD} request can use to reward the rest of participants for their work. This is quite analogous to the way in which miners are rewarded in Bitcoin for their work of including transactions into blocks\cite{bitcoin:paper}. When a \textsf{DON} participant is assigned a \textsf{RAD} task, it will get rewarded with a fraction of the tokens attached to the attestation request if and only if its claim matches the claims brought by a majority of the rest of participants. 
	
	To reduce the computational, energetic and monetary cost of performing \textsf{RAD} tasks, a \textsf{DON} can include some kind of sharding feature that makes it possible to assign a \textsf{RAD} task only to a fraction of the participants.
	
	Given that smart contracts can rely on a \textsf{DON} to decide the outcome of business transactions, it is to be expected that the \textsf{DON} participants may have possible conflicts of interest when performing \textsf{RAD} tasks. In anticipation to this, a \textsf{DON} can implement a reputation scoring system that give participants different chances of being assigned tasks depending on their past performance in terms of honesty.
	
	In addition, if a \textsf{DON} participant decides to tamper with the \textsf{RAD} tasks and bring false, biased or completely made up claims\footnote{To accomplish such a deception, the deceiver would need to modify the code of the \textsf{DON} software he is running in such a way that human intervention would be possible on his own nodes, bypassing the automated nature of the software and breaking the consensus rules of the protocol implicit in the network. The main two reasons for this misbehavior could be conflict of interest (bringing false claims as a response to \textsf{RAD} requests whose outcome could have a potential off-the-network impact on his income) and a mix of sluggishness and greed (based on the assumption that performing \textsf{RAD} tasks entails some kind of marginal cost, a \textsf{DON} participant may choose to not actually perform them and just make the claims up).}, the chances are that such claims will contradict those brought by a majority of protocol-abiding participants, and therefore it will miss the opportunity to collect the token rewards.

	\newpage
	\section{Witnet as a Decentralized Oracle Network}
	\label{sec:Witnet}
	
	The network we are proposing in this work not only satisfies all requirements of a Decentralized Oracle Network (\textsf{DON}) but also applies a series of novel algorithms and strategies to guarantee the quality of the result of \textsf{RAD} requests and reduce if not completely eliminate the chance of manipulation by detecting and penalizing collusion.
	
	Apart from the incentives that the use of token rewards create for participants to behave and not to feed false claims into the \textsf{DON}, we are applying a series of techniques that are heavily influenced by:
	
	\begin{itemize}
		\item Sztorc consensus algorithm, originally presented in the Truthcoin whitepaper\cite{truthcoin}.
		\item Augur consensus algorithm, which in turn is based on Sztorc’s and presented in the Augur whitepaper\cite{augur}.
		\item SchellingCoin conceptual protocol by Buterin\cite{schellingcoin}.
		\item Classic literature on focal points and spontaneous answers in absence of communication. Best example is T. Schelling’s \textit{"The Strategy of Conflict"}\cite{schelling}.
		\item Studies on Reputation-Based voting and how they are affected by collusion attacks, like Bendahmane et.al.\cite{bendahmane}, Watanabe et.al.\cite{watanabe}, Damiani et.al.\cite{damiani} and Xiong et.al.\cite{xiong}.
		\item Studies on collusion detection and design of collusion-resistant protocols and networks, like Porter\cite{porter}, Lepinski et.al.\cite{lepinski1,lepinski2} and Shareef\cite{shareef}.
		\item "Maximum independent set" algorithms for introducing diversity of interests in witnesses selection, like Araujo et.al.\cite{araujo}.
	\end{itemize}
	
	This kind of measures seriously increase the risk inherent to lying because they affect negatively to the cheaters not only by depriving them from short and mid term rewards, but also by denying them the chance to have a say in future tasks which may be more relevant to their interest than the one that caused them to be penalized in the first place.
	
	For this type of incentives and measures to work as intended, it is often necessary that participants of the network (1) have no way to identify each other, (2) have no way to communicate with each other and (3) can not prove the content of their claims to others before the reputation and reward redistribution is made.
	
	As explained by Buterin\cite{schellingcoin}, it could be relatively easy for a single entity controlling something near a 49\% of the \textsf{DON} participants to pre-announce that it will vote for some false claim, and others will also go with that claim out of fear that everyone else will, and if they don not, they will be left out. 
	
	Indeed, with a slightly more twisted scheme, the malicious entity would not even need to participate in the \textsf{DON} at all. The effect is just the same if the entity pre-announces the vote for some false claim, promises a bribe to whoever votes for the same, such bribe is greater than the reward they get from telling the truth, and a majority of participants take the bribe.
	
	However, our protocol renders this kind of gambits completely useless by not giving participants the chance to reveal or prove the actual value of the claims they vote for.
	
	Even if a participant accepts a bribe, it can still tell the truth to the \textsf{DON}, lie to the briber and earn both the reward and the bribe. This is the most profitable of its choices, and therefore the most likely to occur.
	
	For its part, the briber gets zero guarantee that the participants will hold true to their word. Indeed, as the participants are incentivized to act for the good of the \textsf{DON}, tell the truth and deceive the briber, the most likely outcome is that none of the bribed participants will actually vote for the false claim. In a nutshell, any kind of bribe attempt will lead the briber to pour its own resources down the drain.
	
	Witnet, being a tokenized \textsf{DON}, implements also a public ledger that keeps a historic record of all the transactions happening in the network.
	
	Witnet transactions are very similar to Bitcoin transactions in the sense that they unlock tokens from the outputs of previous transactions, aggregate their value, and finally redistribute and lock the value into a new set of outputs.
	
	Transaction outputs use a stack-based scripting language to establish their own spending conditions. This language (explained in more detail in section~\ref{subsec:tx_outputs}) is heavily inspired by Bitcoin's \textsf{Script}, although notable differences exist.
	
	Just like in many other public blockchains, transactions are periodically aggregated into blocks by miners. However, unlike in most of them, the requisite for a Witnet miner to be entitled to \textit{"find a block"} (create a new block that is valid according to the protocol) does not depend on the miner solving a mathematical puzzle. Instead, block miners are chosen by a deterministic algorithm that assigns such roles to participants based on their reputation. The higher their reputation, the more likely they will be elected as block miners. As block miners can collect transaction fees, all participants are incentivized to abide by the protocol, honestly perform the \textsf{RAD} tasks they are assigned and look after their reputation.
	
	In Witnet, the time between creation of new blocks does not depend on a probabilistic process (the time spent by the fastest miner on solving the \textit{proof-of-work} challenge). Instead, it is a deterministic routine---blocks are created periodically, regardless of them being totally full or completely empty.
	
	To successfully bootstrap the Witnet \textsf{DON} and encourage its early adoption, block miners will be also rewarded with a certain amount of new tokens that are freshly created in every block. These \textit{block rewards} are the only way for issuance of new \textsf{Wit} tokens, and their amount will decay over time according to the decreasing-supply algorithm described in section~\vref{subsubsec:wit}.
	
	Witnet focuses on retrieval, attestation and delivery of web contents. Any content that can be publicly accessed through HTTP or HTTPS can also be retrieved, attested and delivered by Witnet. However, other protocols may also be added in the future (FTP, FTPS, SFTP, TFTP, WebDav, BitTorrent, IPFS, etc).
	
	A considerable part of the modern web is no longer static. When you visit a website using your favorite browser, what you see is not only one styled HTML document but the result of many client-side JavaScript computations that alter the document in many ways and even load other documents and asynchronous content in the background. Because of this dynamism, the software in charge of performing web contents retrieval needs to be capable of interpreting websites exactly in the same way that the typical web browser would. At the same time, to cover some very interesting use cases that we explain in section~\ref{subsec:rad_capabilities}, we need such software to be capable of running predefined scripts inside the same web context.
	
	With regards to all these considerations, Witnet miners will use a scriptable \textit{headless browser} to perform the web contents retrieval. A headless browser is a web browser without a graphical user interface. It provides automated control of a website in an environment similar to popular web browsers, but are executed and controlled programmatically by other software. In our case, miners launch instances of the headless browser, make them navigate to the URL specified in the \textsf{RAD} request, run a series of computations (also specified by the request) and simply close the browser when finished.
	
	\subsection{Setting}
	\label{subsec:setting}
	
	\subsubsection{Participants}
	\label{subsubsec:participants}
	
	Any user can participate in Witnet by running a network node that can act as a \textit{client}, a \textit{witness}, and/or a \textit{bridge}.
	
	\begin{itemize}
		\item \textit{Clients} pay tokens to get web contents retrieved, attested and delivered by \textit{miners} through the \textsf{DON}. The due amount of fee to be paid will be discussed in detail later in section~\ref{subsec:fees}---for now, suffice it to say that it depends on the complexity of the \textsf{Retrieve} task, the desired number of \textit{witnesses} to hire for the task and the usage of \textit{bridges}.
		\item \textit{Witnesses} earn tokens in exchange of fulfilling the \textsf{Retrieve} and \textsf{Attest} part of the \textsf{RAD} tasks. Witnessing nodes are also eligible to mine new blocks, and in doing so they hence receive the mining reward for creating a block and transaction fees for the transactions included in the block.
		\item \textit{Bridges} are nodes who specialize in fulfilling the \textsf{Deliver} part of the the \textsf{RAD} tasks. They  are connected to other blockchains and earn tokens by (1) watching other blockchains in search for potential \textsf{RAD} requests to be introduced into Witnet, and (2) replicating the result of the \textsf{RAD} tasks into other blockchains upon request of the \textit{clients}.
	\end{itemize}
	
	\begin{figure}[H]
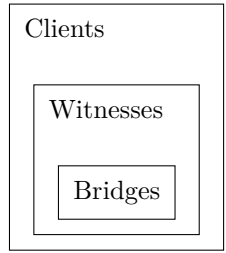

		\centering
		\TBox{
			Clients \\\\
			\TBox{
				Witnesses\\\\
				\TBox{Bridges}
			}
		}
		\caption[Network Participants Subsets]{\textbf{Network participants subsets}}
		\label{fig:participants}
	\end{figure}
	
	Bridges are a subset of the witnesses, which are in turn a subset of all the network participants (the clients).
	
	All witnesses are clients, but not every client needs to be a witness. All bridges are witnesses, but not every witness needs to be a bridge.
	
	\subsubsection{The Network, $\mathcal{N}$}
	\label{subsubsec:network}
	We personify all the users that run Witnet nodes as one single abstract entity: \textit{the network}. The Network acts as an intermediary that runs the \textsf{Manage} protocol at every new block in the Witnet blockchain.
	
	When we refer to network participants as \textit{witnesses}, \textit{miners}, or simply \textit{the network}, we are not meaning actual human beings sitting in front of a computer, fulfilling assignments coming from an Internet overlord that commands them to use their web browser to navigate to a certain website and take a snapshot or copy some text that they must report back. Instead, the participants are computers running a \textit{headless browser} software that automatically receives and executes the \textsf{RAD} tasks without the owner of the computer having to actively do anything else than install such software and configure how much of the available CPU and Internet connection’s bandwidth will be devoted to the \textsf{RAD} tasks.
	
	\subsubsection{The Ledger, $\mathcal{L}$}
	\label{subsubsec:ledger}
	Our protocol is applied on top of an append-only ledger. For generality, we refer to this as \textit{the ledger}, $\mathcal{L}$
	
	Checkpoints are created at regular intervals, and the time between checkpoints is referred to as \textit{epoch}. At any given epoch $t$, all users have access to $\mathcal{L}_t$, a snapshot of the ledger at the preceding checkpoint $t$. Epoch $t$ begins at checkpoint $t$ and finishes at checkpoint $t+1$. This is, checkpoint $t$ closes epoch $t-1$ and opens epoch $t$.
	
	The duration of each Witnet epoch (and thus the time elapsed between checkpoints) is fixed at 90 seconds.
	
	\begin{figure}[H]
		\centering{
			\resizebox{.7\textwidth}{!}{\input{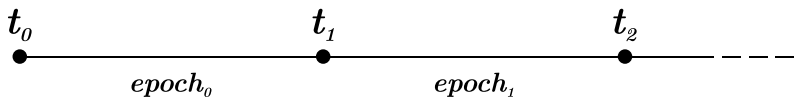}}}
		\caption[Checkpoints and Epochs]{\textbf{Checkpoints and epochs}. Each epoch's number matches the preceding checkpoint.} \label{fig:epochs}
	\end{figure}
	
	There is not a minimum or maximum number of transactions that need to be included in a block. Miners can choose to include as many or as few transactions to the blocks they mine, although they are highly incentivized to include as many as possible in order to collect their fees. Empty blocks (those which include 0 transactions) can also exist. 
	
	Unlike Bitcoin's or Ethereum's blockchains, Witnet's ledger is not a linear chain but a more general form of Directed Acyclic Graph\footnote{In a \textsf{DAG}, every block is linked to any number of blocks, but there is no way to keep following the links from one block through other blocks back to the starting block. This is, no loops are allowed.} (\textsf{DAG}). This means that, for each epoch, more than one block can exist at the same time.
	
	\begin{figure}[H]
		\centering{
			\resizebox{.55\textwidth}{!}{\input{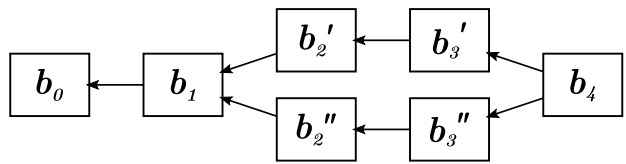}}}
		\caption[Directed Acyclic Graph]{\textbf{Directed Acyclic Graph}. Our protocol uses a type of \textsf{DAG} that allows merging forked yet valid chains without endangering double-spend protection.} \label{fig:dag}
	\end{figure}
	
	For every epoch, a new subset of the network witnesses are pseudo-randomly elected to be the \textit{"block miners"} by the \textit{Reputation-Based Mining Protocol} described in section~\ref{subsec:mining}. Each block miner has the exclusive power to mine (produce and broadcast) exactly one block during its one-epoch "term of office".
	
	Blocks are identified not only for their hash (which is derived from its contents) but also for their checkpoint (which is equivalent to Bitcoin's block height). Two or more blocks for the same epoch can coexist as long as (1) they are correctly formed, and (2) the miners provide valid proofs of their epoch leadership inside the blocks' headers.
	
	Given two or more blocks for the same checkpoint $t$, they will contain a similar (if not the same) set of transactions that were broadcast to the network during epoch $t-1$. Transaction duplicity is addressed by our protocol in the most straightforward way possible: 
	
	\begin{itemize}
		\item Transactions exist in the ledger as independent objects, identified by their hash.
		\item Blocks do not contain transactions per se but \textit{pointers to transactions}.
		\item Two or more blocks for the same checkpoint can contain pointers to the same transaction.
		\item Two or more blocks for different checkpoints can NOT contain pointers to the same transaction. This is, a transaction can not be included into blocks for different checkpoints. The only valid transaction pointer will be the one in the block with the lowest checkpoint.
	\end{itemize}
	
	The double-spend\footnote{Double-spending is an error in a digital cash scheme in which the same single digital token is spent more than once. (Wikipedia)} problem is addressed in a similar way:
	
	\begin{itemize}
		\item Just like in Bitcoin, Witnet transactions unlock existing UTXO\footnote{Unspent Transaction Output.}s, redistribute the unlocked coins and lock them in a new set of UTXOs.
		\item The value of Witnet outputs is not expressed as an absolute number of coins but as a percentage of the sum of the values of the inputs.
		\item A number $n$ of transactions spending the same UTXO with value $v$ in the same block or in two different blocks for the same checkpoint are perfectly valid, but the value is equally split among their coincident inputs. This is, each of the $n$ transactions will be able to spend at most $\frac{v}{n}$ coins.
		\item Witnet transactions have a series of special properties that make them trivial to be efficiently processed in parallel. Those properties are put forward in section~\vref{subsec:tx_properties}.
	\end{itemize}
	
	\subsubsection{The Coin, \textsf{Wit}}
	\label{subsubsec:wit}
	
	\textsf{Wit} is Witnet's native token. Its generation algorithm defines, in advance, how new coins (token units) will be created and at what rate. Any coin that is generated by a malicious miner that does not follow the rules will be rejected by the network and thus is worthless.
	
	\textsf{Wit}s are created every time a miner publishes a new block. This is called the \textit{"block reward"}. In the event that two or more blocks were added to the ledger at the same time, the amount of coins created would be divided between the miners who mined them.
	
	The number of \textsf{wit}s generated per block starts at 500 and is set to decrease geometrically, with a 50\% reduction every 1,750,000 blocks, or approximately 5 years. Each of these periodic reductions is known as \textit{halving}. The result is that the number of \textsf{wits} ever created by the issuance mechanism will never exceed 2,500,000,000.
	
	\begin{figure}[H]
		\[ \sum_{i=0}^{\infty} \frac{500 \cdot 1750000}{2^i} \approx 2500000000\]
		\caption[Controlled Supply Generation Algorithm]{\textbf{Controlled supply generation algorithm}. As time passes, the issuance rate is periodically halved and the amount of coins ever issued by the mining rewards mechanism approaches the 2,500,000,000 limit. In this model, $i$ represents the number of halving events ever passed.}
		\label{fig:issuance}
	\end{figure}
	
	\begin{figure}[H]
		\centering
		\parbox{7cm}{
			\centering
			\begin{tikzpicture}
			\begin{axis}[
			axis lines=left,
			ticks=none,
			ymax=20000000,
			xmax=20,
			height=4cm
			]
			\addplot[color=red, domain=0:20, samples=50]{(500*1750000)/(2^x)};
			\end{axis}
			\end{tikzpicture}
			\caption[Coin Issuance Rate]{Coin issuance rate (inflation) will rapidly decay over time. Note that after only a few reward ages, block rewards will be marginal.}
			\label{fig:aggregationA}}
		\qquad
		\begin{minipage}{7cm}
			\centering
			\begin{tikzpicture}
			\begin{axis}[
			axis lines=left,
			ticks=none,
			ymin=0,
			ymax=3000000000,
			xmax=20,
			height=4cm
			]
			\addplot[color=blue, domain=0:20, samples=50]{(750000000+13671875*2^(7-x)*(2^x-1))};
			\end{axis}
			\end{tikzpicture}
			\caption[Total Coin Supply]{The total number of coins in circulation will rapidly grow at the beginning but then start to slow down after only a few reward ages.}
			\label{fig:aggregationB}
		\end{minipage}
	\end{figure}
	
	The inflation rate steadily trends downwards. The block reward given to miners is made up of newly-created \textsf{wits} plus transaction fees. As inflation tends to zero over time, miners will obtain an income only from transaction fees, which will provide an incentive to keep mining to make transactions irreversible.
	
	\subsubsection{The Headless Browser, $\mathcal{B}$}
	\label{subsubsec:browser}
	
	As mentioned earlier in the introduction of section~\ref{sec:Witnet}, witnesses use a scriptable \textit{headless browser} to perform the web contents retrieval.
	
	A headless browser is a web browser without a graphical user interface. It provides automated control of a website in an environment similar to popular web browsers, but are executed and controlled programmatically by other software.
	
	In our case, witnessing nodes launch instances of the headless browser, make them navigate to the URL specified in the \textsf{RAD} request, run a series of computations (also specified by the request) and simply close the browser when finished.
	
	\textsf{RAD} requests can contain a small script that tells the headless browser which specific data units to pick from the visited website. That is the case for \textit{Specific Data Units Retrieval}, explained in section~\vref{subsubsec:data_units}.
	
	One \textsf{RAD} request can cause more than one instance of the headless browser to be spawned in parallel. That is the case for \textit{Multiple Sources Fact Cross-Checking}, explained in section~\vref{subsubsec:crosscheck}.
	
	Using a headless browser also makes the witnesses indistinguishable from a human being using a regular web browser to navigate a website, so it prevents an hypothetical kind of attack in which the administrator of a website---aware of his site being used to influence the outcome a smart contract---may tamper with the attestations by presenting fake or contradicting information to miners or even blocking them altogether.
	
	\subsection{Retrieval and Attestation Capabilities}
	\label{subsec:rad_capabilities}
	
	\subsubsection{General Case}
	\label{subsubsec:rad_general_case}
	
	At their simplest, these are the main steps of Witnet's \textsf{RAD} flow:
	
	\begin{enumerate}
		\item Alice wants to get a web content retrieved, attested and delivered to her.
		\item Alice prepares a \textsf{RAD} request that will be sent to the network. She needs to attach a certain amount of tokens to the request. This amount will depend on (1) the complexity of the retrieval, (2) the level of certainty that she expects for the attestation (the number of witnesses that will be employed), and (3) usage of \textsf{Retrieve} clauses in the request. \textsf{RAD} fees are further discussed in section~\ref{subsec:fees}.
		\item Alice encodes the \textsf{RAD} request as a Witnet transaction and send it to the network.
		\item The task of solving Alice's request is assigned to a subset of all the witnesses. These are elected by consensus thanks to the \textit{Reputation-Based Task Assignment Protocol} described in section~\ref{subsec:assignment}. Each of these miners will:
		\begin{enumerate}
			\item Retrieve the web contents that are specified in the request using the headless browser as described in section~\ref{subsubsec:browser}. The resulting value is what we call a \textit{claim}.
			\item Calculate a \textit{nonced hash} of the claim. This hash will be different for each witness and will be derived from the claim itself, the miner's public key and the hash of the latest block.
			\item Send the hash of the claim to the network as a commitment to publish the actual claim when the rest of designated witnesses have also made their own commitments. The witnesses also use these commitment transactions as a pledge for reserving for themselves a share of the tokens attached to the \textsf{RAD} request.
		\end{enumerate}
		\item Once all the designated witnesses have made their pledges, the block miner(s) will divide the reward tokens that are attached to the \textsf{RAD} request among all the witnesses who made a valid pledge for such request.
		\item The witnesses who made their pledges will start to reveal the actual retrieved contents. They will do so by sending \textsf{reveal-redeem} transactions that spend their commitment transactions and send the reward tokens to their own wallet.
		\item The block miner(s) for the next block will have to compare the claims coming from different witnesses and choose which is the winning claim (nominally, \textit{"the truth"}) by applying the \textit{Truth-By-Consensus} algorithm explained in section~\ref{subsec:svd}. 
		\begin{itemize}
			\item The witnesses who told the truth will earn reputation points and the transactions redeeming their share of the reward will be accepted and included into next block.
			\item On the contrary, those witnesses who lied will lose a fraction of their reputation points and they will not be able to redeem their share of the reward.
		\end{itemize}
		\item At this point, the result of the attestation will be public and available to Alice as well as to any other participant of the network.
		\item If a \textsf{Deliver} clause was specified in the request, bridges\footnote{See section~\vref{subsubsec:participants}.} capable of performing such type of tasks will come into play and do their job (e.g.: \textit{"Send the result as a parameter to the \textsf{callback()} function in the Ethereum smart contract with address 0xe711fA745e..."}).
	\end{enumerate}
	
	\begin{figure}[H]\centering
		\[ \left.\begin{array}{cc}
		Claim_{witness1} \\ 
		Claim_{witness2} \\ 
		Claim_{witness3}
		\end{array}\right\}\rightarrow \textsf{Truth-By-Consensus} \rightarrow \textbf{Result}\]
		\caption[Truth-By-Consensus]{\textbf{Truth-By-Consensus}. For each \textit{RAD} request, the block miner(s) compare the claims coming from  all the designated wtinesses, weight the values depending on each one's reputation and compute the result using the \textit{Truth-By-Consensus} algorithm as described in section~\ref{subsec:svd}}
		\label{fig:network_flow}
	\end{figure}
	
	Section~\vref{subsec:protocol} defines the client and miner cycles in detail.
	
	There exist some use cases described hereafter that despite of introducing some additional complexity are addressed by the proposed network without any problems.
	
	\subsubsection{Specific Data Units Retrieval}
	\label{subsubsec:data_units}
	For example, Alice might be interested in the acquisition and attestation of a single data unit out of a whole web page. 
	
	For these cases, Witnet provides a flow-based, tacit, point-free scripting language that will let the client to (1) indicate the specific element to be retrieved, attested and delivered, and (2) apply basic transformations on it.
	
	For the sake of ease, such language can be implemented as a \textit{domain specific language} (DSL) on top of the Javascript programming language, which is already known by most smart contracts developers and web developers.
	
	The combination of one URL and one of these scripts is called a \textit{retrieval path}.
	
	The \textit{\textsf{RAD} cost} mentioned earlier in this chapter is directly affected by the computational complexity of the retrieval path being included in the request, as well as by the data size of the returned value. \textsf{RAD} fees are further discussed in section~\ref{subsec:fees}.
	
	\begin{figure}[H]
		\[ Acquisition + Normalization = RetrievalPath\]
		\caption[Retrieval Paths]{\textbf{Retrieval paths}. Each retrieval path is composed of one acquisition (URL) and one normalization script.}
		\label{fig:nodes_flow}
	\end{figure}
	
	\subsubsection{Multiple Sources Fact Cross-checking}
	\label{subsubsec:crosscheck}
	
	Alice may also want to cross-check some fact by requesting the retrieval and attestation of data units representing the same reality but published by different entities across a number of web sites.
	
	Our protocol covers this case as well by allowing the client to include multiple retrieval paths in a single attestation request. In addition, the scripting language introduced in section~\vref{subsubsec:data_units} includes normalization and aggregation methods in a MapReduce style that enable the merging of data coming from a plurality of sources despite of the slight format differences that might exist between them.
	
	Generally speaking, a \textsf{RAD} request must contain one or more retrieval paths and the definition of one \textit{path aggregation function}.
	
	For every retrieval path, a headless browser instance is launched, the corresponding URL is loaded and the matching normalization script is run. Once the normalization scripts from all the retrieval paths have finished running, an additional instance of the headless browser is created and pointed to a clean context in which the aggregation function is run over a list containing the results of each retrieval path.
	
	The whole retrieval flow performed by each miner in a multiple sources fact cross-checking \textsf{RAD} request is depicted in Figure~\vref{fig:nodes_flow}.
	
	\begin{figure}[H]\centering
		\[ \left.\begin{array}{cc}
		Acquisition_{a} + Normalization_{a} = RetrievalPath_{a} \\ 
		Acquisition_{b} + Normalization_{b} = RetrievalPath_{b} \\ 
		Acquisition_{c} + Normalization_{c} = RetrievalPath_{c}
		\end{array}\right\} \to Aggregation \to \textbf{Claim}\]
		\caption[Content Retrieval Flow]{\textbf{Content retrieval flow}. Each designated witness brings a single claim that is derived from the aggregation of the values that result after applying normalization methods on the acquired web contents.}
		\label{fig:nodes_flow}
	\end{figure}

	\subsubsection{Future Facts}
	\label{subsubsec:future}
	
	It is also conceivable that Alice may want to retrieve and attest a verified piece of data whose value is unknown, uncertain or impossible to resolve at the time of formulation of a \textsf{RAD} request.
	
	\begin{example}[example:future]
		A smart contract is set to have a different outcome depending on the answer to the question "\textit{How much will 1 bitcoin be worth in 2 years from now?}".
		\tcblower
		Such outcome can not be trivially resolved today because of indetermination: the answer is yet unknown and will only be known once the date stated by the question itself has come.
	\end{example}
	
	The inability to evaluate this kind of questions comes from the statements having a verification precondition---either expressed or implied---that must be met before the claim itself can be evaluated and verified. In the Example~\vref{example:future}, the precondition is \textit{"Has it been 2 years now since the request was formulated?"}.
	
	Although the kind of functionality implemented by a \textsf{DON} is commonly known as \textit{oracle}, Witnet is radically different to the oracles from the classical antiquity in the sense that we make an important distinction between predictions or beliefs (claims that \textit{may} hold true but are impossible to verify) and truth (claims that are verifiable at this time).
	
	As we the ordinary mortals have no means to verify a claim whose veracity is not yet determinable or to foresee the future answer to simple questions like the one in Example~\vref{example:future}, the only thing we can do is waiting for such claims to eventually become verifiable.
	
	For these cases, the proposed network includes a \textit{time lock} feature that, upon request, will keep a \textsf{RAD} request unresolved until a certain point in the future. Once that point in time has passed, \textit{witnesses} are strongly incentivized to immediately resolve the request in order to harvest its fees. In anticipation that witness fees and block miner fees may grow over time, a \textit{Replace-By-Fee} feature will allow a client to "republish" its request using an increased reward as long as the retrieval paths remain the same.
	
	\subsection{Retrieval and Attestation Limitations}
	\label{subsec:limitations}
	
	\subsubsection{Undecidability}
	\label{subsubsec:undecidability}
	
	There exist claims and statements whose truth or falsehood are neither provable nor refutable. These are called \textit{undecidable statements}.
	
	Undecidable statements and undecidable problems have been object of abundant study by many authors during the last century, notably Gödel\cite{godel},  Church\cite{church}, Turing\cite{turing}, Rosser\cite{rosser}, Rice\cite{rice}, Kleene\cite{kleene} and Conway\cite{conway}.
	
	The concept of undecidable statements is itself based in the notion of \textit{decidability}. In logic, decidability is the question of the existence of an effective method to determine the truth or falsehood of a statement.
	
	\begin{itemize}
		\item A statement whose truth or falsehood can be evaluated in the present is decidable.
		\item A statement whose truth or falsehood can not be evaluated in the present but will become assessable in the future is also decidable.
		\item However, a statement whose truth or falsehood can not be evaluated to a correct (determined and well-formed) value after finite though possibly long time, is considered undecidable.
	\end{itemize}
	
	The problem behind establishing the decidability of a certain statement is quite equivalent to the halting problem in computability theory: determining, from a description of a computer program and an input, whether the program will eventually finish running and return a valid output or whether on the contrary it will continue to run forever.
	
	Just like in the halting problem, which was proved undecidable by Alan Turing in 1936\cite{turing}, the assessability of a certain statement is itself undecidable. Although common sense can tell us in most cases whether a statement is assessable at the moment of formulation or whether it will be assessable in the near future, there still exist statements whose assessability can not be predicted or assured to the 100\%.
	
	\begin{example}[example:undecidable]
		A smart contract is set to have a different outcome depending on the answer to the question "\textit{Will 1 bitcoin ever be worth \$50K?}".
		\tcblower
		Such outcome can not be trivially resolved today not only because of indetermination---the answer is yet unknown---but also because of undecidability, as we can not foresee when will the question itself be answerable or whether it will ever have an answer whatsoever.
	\end{example}
	
	Like with the question in the Example~\ref{example:future} in section~\ref{subsubsec:future}, the inability to evaluate this kind of questions also comes from the statements having a verification precondition---either expressed or implied---that must be met before the statement itself can be evaluated and verified. In the Example~\vref{example:undecidable}, the precondition is \textit{"Has 1 bitcoin ever been worth \$50K?"}\footnote{Note that in the given example the check in the precondition is the same as the question itself but in past tense. This is a special case in which the statement can never be evaluated to a \textsf{false} value: before the precondition is met the result is undetermined, and after it is met the result is always \textsf{true}.}.
	
	For these cases, the Witnet protocol specifies a special type of \textsf{RAD} requests that remain resolved indefinitely. This is, no miner is able to pledge a solution for it.
	
	However, if an undecidable request eventually becomes decidable, the client can produce a transaction that will "relaunch" the request, but this time as a regular \textsf{RAD} request so that miners can start pledging solutions to it as soon as the next epoch starts.
	
	\subsubsection{Unverifiability}
	\label{subsubsec:verifiability}
	
	As seen in sections \ref{subsubsec:future} and \ref{subsubsec:undecidability}, not all claims are verifiable, being indetermination and undecidability the two main reasons for such lack thereof.
	
	Verifiability or provability is the capability of a certain statement to be demonstrated, verified, confirmed, substantiated or logically proved.
	
	Douglas R. Hofstadter, in his 1979 Pullitzer-winning book \textit{"Gödel, Echer, Bach: An Eternal Golden Braid"}, states that \textit{"provability is a weaker notion than truth"}\cite{hofstadter}. We often can not prove things that we know are true.
	
	As regards Witnet, the only possible truth is the verifiable one. Indeed, verifiability is hard-coded into the design of the network. The mere act of performing distributed retrieval of data is in itself a form of verification, specially if several web sources are queried via multiple acquisition paths as described in sections \ref{subsubsec:data_units} and \vref{subsubsec:crosscheck}. In the same manner, the fact that a final agreed claim emerges from the aggregation of all the claims brought by a plurality of miners proves itself that the claim was verifiable at the time of the attestation.
	
	There exist, of course, realities and facts that, despite of their truthness, can not be processed as true statements by Witnet because of their lack of verifiability. It is not that the network will fail to resolve a request expressing a question whose answer is unverifiable. As a matter of fact, such type of questions is just impossible to translate into \textit{RAD} requests. This is because these requests---and more specifically \textit{retrieval paths} as described in section \vref{subsubsec:data_units}---are explicit about how claims are extracted and derived from available information online. That is why the only impossible attestation is the one with retrieval paths pointing to nonexistent web contents or applying transformations on the retrieved data in such a way that different paths from a single request result in contradicting claims. In other words, the success of a \textsf{RAD} request depends exclusively on its own design, so it is the responsibility of the client to only include valid and provable retrieval paths.
	
	\subsection{The Protocol}
	\label{subsec:protocol}
	
	In this section, we give an overview of the Witnet \textsf{DON} by describing the operations performed by the clients, the Network and the different types of miners. 
	
	\subsubsection{\textit{Client} Cycle}
	\label{subsubsec:client_cycle}
	
	Clients performing \textsf{RAD} requests run the following protocols in addition to the ones described in the \textit{network} Cycle in section~\vref{subsubsec:network_cycle}.
	
	\begin{enumerate}
		\item \textbf{RAD-Post}: \textit{Client requests the retrieval, attestation and delivery of web contents.}
		
		Clients can request the retrieval, attestation and delivery of web contents by paying witnesses in \textsf{Wit} tokens.
		
		A client initiates the \textsf{RAD-Post} protocol by submitting a \textsf{RAD} client \textsf{request} transaction to the network. This type of transaction is further described in section~\vref{subsec:tx_types}.
		
		Clients can decide the amount of miners that will be assigned to each \textsf{RAD} task by specifying a replication factor in the request. The minimum replication factor is 2. Higher redundancy results in a higher certainty and confidence of the attestation. A replication factor around 6 should be more than enough for most cases while keeping costs under control\footnote{A maximum replication factor shall also be imposed in order to avoid abuse of the \textsf{DON} to conduct (expensive) DDoS-attack on websites. In the same manner, the network should only allow a limited number of \textsf{RAD} requests that included equivalent retrieval paths and are set to be resolved at the same checkpoint.}.
		
		The \textsf{RAD} client \textsf{request} transaction must pay a miner fee appropriate to its own size when serialized. This is, the sum of the values of its inputs must exceed that of the outputs in a sufficient amount to make it attractive for block miners to include it into a block.
		
		\item \textbf{RAD-Get}: \textit{Client reads the result of \textsf{RAD} requests.}
		
		In the same moment that block miners apply the \textsf{Truth-By-Consensus} algorithm on the claims brought by witnesses, a final \textit{"verdict"} pointing to the retrieved and attested content emerges. The \textsf{reveal-redeem} transactions containing the value of the retrieved and attested content (\textit{the solution}) are immediately included into the block being mined.
		
		As any client can see and read the ledger at any time, as soon as the block containing the solution is broadcast to the network, it can be easily traced back to the related \textsf{RAD-Post} transaction and matched with the original request.
		
		Therefore, the \textsf{RAD-Get} protocol can be run locally---thus with no transaction cost---by any client with an up-to-date copy of the Witnet ledger.
	\end{enumerate}

	\subsubsection{\textit{Witness} Cycle}
	\label{subsubsec:miner_cycle}
	
	Witnesses perform the following protocols in addition to the ones described in the \textit{network} cycle in section~\vref{subsubsec:network_cycle}.
	
	\begin{enumerate}
		\item \textbf{Receive \textsf{RAD} requests}: \textit{Witnesses read \textsf{RAD-Post} requests from the blockchain.}
		
		From the moment that a \textsf{RAD-Post} gets broadcast to the network, all witnesses have all the information they need to start working on the task by executing its \textsf{Retrieve} part.
		
		However, if the request has a precondition (see \textit{time locked} and \textit{undecidable} requests in sections \ref{subsubsec:future} and \ref{subsubsec:undecidability}), it must be kept unresolved until its precondition is met.
		
		Any \textsf{RAD} request without a precondition can be worked on immediately, although miners should refrain from doing so until they know if they have been designated as witnesses for such task. Otherwise, they may be spending resources in exchange of no reward at all.
		
		\item \textbf{Discover \textsf{Retrieve} tasks assignments}: \textit{Witnesses apply the task assignment protocol to discover task assignments.}
		
		At any moment, a witness can locally run the \textit{Reputation-Based Task Assignment Protocol} described in section~\ref{subsec:assignment} on all the \textsf{RAD} requests broadcast during the current epoch to figure out which \textsf{RAD} tasks it has been designated for.
		
		If there were any pending \textsf{RAD} requests with a time lock precondition expiring in the last checkpoint (see section~\ref{subsubsec:future}), they can also be checked for assignment using the same task assignment protocol.
		
		\item \textbf{\textsf{Retrieve} and \textsf{commit-pledge}}: \textit{Witnesses perform the \textsf{retrieve} part of their assigned \textsf{RAD} tasks and pledge to publish the results in the future.}
		
		At this point, each witness should start fulfilling their assigned \textsf{RAD} tasks using their bundled headless browser as described in section~\ref{subsubsec:browser}. For every \textsf{RAD} task, each designated miner will come up with one and only one claim.
		
		Once they have their claims, witnesses can compose their \textsf{commit-pledge} transactions and broadcast them right away. In these transactions, witnesses must also include all information necessary to prove their task assignment to the rest of the network.
		
		\item \textbf{\textsf{Reveal-redeem}}: \textit{Witnesses reveal the results of their assigned \textsf{RAD} tasks and redeem their reward.}
		
		Once the \textsf{commit-pledge} transactions have been included in a block, each of the witnesses who made the pledges must reveal their claims and provide everything necessary to prove that (1) the nonced hashes that they committed in their respective \textsf{commit-pledge} transactions were actually derived from the revealed claims, and (2) they are the authors of the pledges they are trying to redeem.
		
	\end{enumerate}
	
	\subsubsection{\textit{bridge} Cycle}
	\label{subsubsec:delivery_cycle}
	
	Bridge nodes (as defined in section~\ref{subsec:bridges}) are in charge of delivering the claims that result from \textsf{Request-Attest} tasks to other blockchains. Bridges also monitor those other blockchains in search for potential \textsf{RAD} requests to be introduced into Witnet.
	
	They perform the following protocols in addition to the ones described in the \textit{witness} cycle in section~\vref{subsubsec:miner_cycle} and the \textit{network} cycle in section~\vref{subsubsec:network_cycle}.

	\begin{enumerate}
		\item \textbf{Discover \textsf{Deliver} task assignments}: \textit{Bridges apply the task assignment protocol to discover \textsf{Deliver} task assignments.}
		
		At any moment, a bridge node can locally run the \textit{Reputation-Based Task Assignment Protocol} described in section~\ref{subsec:assignment} on all the \textsf{RAD} requests resolved during the current epoch to figure out which \textsf{Delivery} tasks it has been assigned to.
		
		\item \textbf{\textsf{Deliver} attested claims}: \textit{Bridges perform delivery of attested claims to other blockchains.}
		
		Bridges are in charge of reporting the results of \textsf{RAD} tasks to other blockchains. In exchange of performing this work and spending their own time and coins in sending their reports, they will be awarded with \textsf{Wit} tokens that were allocated for such purpose in the request transaction.
		
		\item \textbf{Discover outer \textsf{RAD} requests}: \textit{Bridges discover \textsf{RAD} requests posted in other blockchains.}
		
		Bridges monitor other blockchains in search of \textsf{RAD} requests codified inside transactions broadcasted by unkown cients in those blockchains. When bridges find one of these transactions, they read its payload and convert it into a valid Witet \textsf{RAD} request.
		
		\item \textbf{\textsf{RAD-Post}}: \textit{Bridges post \textsf{RAD} requests in behalf of users of other blockchains.}
		
		Bridges must broadcast the \textsf{RAD} request they have built in behalf of the unkown client from the outer blockchain. In exchange of performing this work and spending their own \textsf{Wit}s, bridges will need to be rewarded by the unknown clients in whatever form they accept (likely in the native tokens of the outer blockchain).
		
	\end{enumerate}
	
	\subsubsection{\textit{Network} Cycle}
	\label{subsubsec:network_cycle}
	
	For every Witnet epoch, the network (all participants) must:
	
	\begin{enumerate}
		\item \textbf{Receive and validate last epoch's block(s)}:
		
		At the start of each epoch, the network checks that all the blocks received for the last epoch contain a valid proof of leadership and a commitment (signed reference) to one or more blocks from the previous block.
		
		Each participant of the network must check if the transactions included by the miners into the blocks have been previously received and checked for validity. If not, they must request their peers for the newly discovered transactions.
		
		The network must then create a Merkle tree with all the transactions included in each of the blocks and check if the root of the resulting tree matches the root used in the block header.
		
		\item \textbf{Receive and validate incoming transactions}:
		
		During each epoch, all the transactions broadcast by the clients are sent to each of the participants of the network. Upon reception of a new transaction, the network must check its validity (correctness of their inputs and success of redeem scripts) and store it for posterior inclusion into blocks.
		
		\item \textbf{Check for epoch leadership}:
		
		At every epoch, all network participants (clients, witnesses and bridges) can use the \textit{Reputation-Based Mining Protocol} from section~\ref{subsec:mining} to discover if they have been elected block miners for the current epoch. If so, they will be implicitly authorized by the network to produce and broadcast exactly one block once the epoch has finished.
		
		\item \textbf{Mine a block}:
		
		At each epoch checkpoint, every block miner has the power to mine (produce and broadcast) one block.
		
		Block miners can include as many transactions as they want into a single block, with the only limitation being keeping the size of the block under 1MB. Predictably, block miners will prioritize small transactions paying high rewards over bigger transactions paying lower rewards\footnote{Note that \textit{small} and \textit{big} are not referring to the amount of tokens being transferred but to the size in bytes of the transaction when serialized as described in section~\ref{subsec:serialization}.}.
		
		A block miner who created a new block needs to prove its right to do so to the rest of the network by including a proof in the block header. These "proofs of leadership" are described in detail in section~\ref{subsec:mining}.
		
		Each of the block miners who created a new block must broadcast their blocks to the rest of the network during the next epoch. Otherwise, their blocks may be rejected by the network.
		
	\end{enumerate}
	
	\newpage
	\section{Reputation}
	\label{sec:reputation}
	
	A key feature of Witnet is \textit{reputation points}. The total amount of reputation points in the system is a fixed quantity, determined upon the launch of Witnet. Holding reputation points entitles a Witnet witness to be elected for performing \textsf{RAD} tasks and more generally any network participant to mine mine new blocks. The higher the reputation score a participant has, the more likely it will be able to collect token rewards from performing \textsf{RAD} tasks and mining new blocks.
	
	Witnet reputation points work in a similar way to Augur's \textit{REP}\cite{augur}, Truthcoin's \textit{Votecoins}\cite{truthcoin} and, to a lesser extent, Filecoin's \textit{Power}\cite{filecoin}---they are gained and lost depending on how reliably their owner votes with the consensus.
	
	Reputation points clearly define and limit the fuzzy concept of "reputation". The existence of a fixed amount of total reputation points provides Sybil attack immunity\footnote{The proposed network is designed to not depend in any way on the number of participants in the network. All of its implicit economic models work just the same regardless of whether there is a single large actor holding a big part of the reputation or a million small actors holding the same amount of reputation points.} and at the same time gives the network an effective way to penalize miners for laziness.
	
	Reputation points are affected by \textit{demurrage}\footnote{Demurrage is the cost associated with owning or holding currency over a given period. It is sometimes referred to as a carrying cost of money. For commodity money such as gold, demurrage is the cost of storing and securing the gold. For paper currency, it can take the form of a periodic tax, such as a stamp tax, on currency holdings.---\href{https://en.wikipedia.org/wiki/Demurrage_(currency)}{Wikipedia -- Demurrage (currency)}.}: network participants lose reputation if they hoard their points instead of using them to become witnesses and have a say in the outcome of \textsf{RAD} requests. In that sense, it can be said that reputation points are a liability as well as an asset, because their owners are obliged to put them to proper use or lose them altogether if they do not. On the contrary, \textsf{Wit}---Witnet's native token---is not affected by this demurrage policy.
	
	\subsection{Reputation Protocol}
	\label{subsec:reputation_protocol}
	
	\begin{itemize}
		\item Every possible public key has its own reputation score\footnote{Witnet uses \textsf{ECDSA} over the \textsf{secp256k1} curve as its main signature algorithm. This setup can accommodate up to $2^{256}$ different public keys.}.
		\item At checkpoint \textsf{0} of the ledger, all public keys will have equal reputation scores of 1.
		\item The sum of the scores of all the public keys is fixed and will equal $2^{256}$ points at all times.
		\item New reputation points can not be created after checkpoint \textsf{0}. 
		\item Reputation points can not be destroyed and they never leave the network.
		\item Reputation points are earned by witnesses when they agree with a majority of the rest of the designated witnesses on the resulting claims of the \textsf{RAD} tasks they get assigned.
		\item Reputation points are lost by witnesses when they contradict or fail to agree with a majority of the rest of the miners on the resulting claims of the \textsf{RAD} tasks they get assigned.
		\item At every epoch checkpoint, reputation points are lost by all the network participants at once as a form of demurrage. Those points are rewarded to the honest miners that fulfilled such epoch's \textsf{RAD} tasks.
		\item At every epoch, the sum of reputation points earned by honest miners equals the sum of points deducted from dishonest ones plus the sum of points deducted from every participant by the demurrage system.
		\item If during a certain epoch every witness is honest, none of them will lose reputation. However, demurrage will still apply to all participants, and the deducted points will be distributed evenly among the honest witnesses and bridges that fulfilled such epoch's \textsf{RAD} tasks.
	\end{itemize}
	
	In Witnet, being elected for block mining or designated for fulfilling \textsf{RAD} tasks works just like a lottery in which reputation points are the lottery tickets. The more reputation a participant owns, the bigger its chances to win the right to collect block and tasks rewards.
	
	To ensure that the most reputable participants also bear a greater liability, the reputation demurrage function has been modeled in such a way that it applies a deduction on each reputation score in proportion to the score itself. It causes the reputation score of the most reputable participants to decay rapidly while the score of the smallest participants is left almost intact.
	
	This type of progressive demurrage can also be seen as a measure to fight concentration and favor redistribution of reputation points, just like wealth taxes take a larger percentage from high-income earners than they do from low-income individuals.
	
	Figure~\vref{fig:reputation_decay} depicts the function that governs Witnet's reputation demurrage system. The \textsf{decay rate} ($\mathcal{D}$) must lie in the unit interval $[0,1]$, so $0 \leq \mathcal{D} \leq 1$. In the proposed function, this rate is a parameter that can be adjusted to make the reputation system more or less progressive. The closer that $\mathcal{D}$ gets to $0$, the more that the most reputable participants will be urged to participate in the system. On the contrary, the closer $\mathcal{D}$ gets to $1$, the more that they will be able to remain idle without a profound negative impact to their potential income.
	
	\begin{figure}[H]
		\centering
		\[ \mathcal{R}_{epoch_t} = \mathcal{R}_{epoch_{t-1}} \cdot \mathcal{D}^{\log_{10}(\mathcal{R}_{epoch_{t-1}})} \]
		\caption[Reputation Demurrage Function]{\textbf{Reputation demurrage function}. The logarithmic exponentiation of the \textsf{decay rate} ($\mathcal{D}$) causes the score of the participants with the biggest reputation stake to decay more rapidly.}
		\label{fig:reputation_decay}
	\end{figure}
	
	We are proposing an initial decay rate of $\mathcal{D} = 0.99$. Although this rate could seem too conservative at first sight, it applies a reasonable decay to participants of all sizes, as depicted in figure\vref{fig:decay_table}.
	
	\begin{figure}[H]
		\centering
		\begin{tabular}{|llllllllll|}
			\cline{2-10}
			\multicolumn{1}{l|}{}      & \multicolumn{9}{c|}{Epochs}                                              \\
			\multicolumn{1}{l|}{}      & 0    & 1      & 25     & 50     & 75     & 100   & 125   & 150   & 175   \\ \hline
			\multicolumn{1}{|l}{Scores}& 1    & 1      & 1      & 1      & 1      & 1     & 1     & 1     & 1     \\ \cline{2-10} 
			\multicolumn{1}{|l}{}      & 10   & 9.90   & 7.87   & 6.36   & 5.25   & 4.42  & 3.79  & 3.30  & 2.91  \\ \cline{2-10} 
			\multicolumn{1}{|l}{}      & 100  & 98.01  & 62.06  & 40.46  & 27.58  & 19.56 & 14.37 & 10.90 & 8.51  \\ \cline{2-10} 
			\multicolumn{1}{|l}{}      & 1000 & 970.29 & 480.99 & 257.42 & 144.85 & 86.51 & 54.50 & 36.01 & 24.84 \\ \cline{2-10}
			\multicolumn{1}{|l}{}      & 10000& 9605.26& 3851.53& 1637.54& 760.72 & 382.61& 206.63& 118.95& 72.50 \\ \hline
		\end{tabular}\\
		\begin{tabular}{|lllllllllllll|}
			\cline{1-13}
			\multicolumn{13}{|c|}{Epochs}                                                                       \\
			200  & 225  & 250    & 275    & 300    & 325    & 350   & 375   & 400   & 425   & 450   & 475 & 500 \\ \hline
			1    & 1    & 1      & 1      & 1      & 1      & 1     & 1     & 1     & 1     & 1     & 1   & 1   \\ \hline
			2.61 & 2.36  & 2.16  & 1.99   & 1.85   & 1.74   & 1.64  & 1.56  & 1.49  & 1.43  & 1.37  & 1.33& 1.29\\ \hline
			6.82 & 5.59  & 4.67  & 3.98   & 3.45   & 3.03   & 2.70  & 2.44  & 2.22  & 2.04  & 1.90  & 1.77& 1.67\\ \hline
			17.81& 13.21 & 10.11 & 7.96   & 6.42   & 5.29   & 4.45  & 3.81  & 3.32  & 2.93  & 2.62  & 2.37& 2.17\\ \hline
			46.52& 31.25 & 21.87 & 15.89  & 11.93  & 9.23   & 7.33  & 5.96  & 4.95  & 4.20  & 3.61  & 3.16& 2.81\\ \hline
		\end{tabular}
		\caption[Reputation Demurrage Simulation]{\textbf{Reputation demurrage simulation}. In this simulation with $\mathcal{D} = 0.99$, 5 participants remain idle (not accepting \textsf{RAD} tasks) for 500 epochs (12.5 hours). At $epoch_{0}$ the differences among each participant's scores is \textsf{x10}. At $epoch_{200}$, demurrage has reduced such differences to less than \textsf{x3}, and at $epoch_{500}$ their scores have nearly converged. Over the total period, the score for each of the participants was divided by: $1$, $8$, $60$, $461$ and $3559$ respectively.}
		\label{fig:decay_table}
	\end{figure}
	
	In Witnet, reputation has the following properties:
	
	\begin{itemize}
		\item \textit{Public}: By reading the blockchain, anyone can calculate the reputation of each participant at any point in time.
		\item \textit{Verifiable}: Reputation is earned and lost by performing \textsf{RAD} tasks whose outcome is publicly available. By reading the blockchain, anyone can verify the outcome of those tasks and check if the reputation claimed by a participant is correct.
		\item \textit{Fluid}: At any point in time, participants can earn reputation points by becoming witnesses and performing \textsf{RAD} tasks honestly; or lose them if they tamper with those tasks or simply ignore them. In this way, reputation points are always flowing from those who do not contribute to the system toward those who do.
	\end{itemize}
	
	\subsection{Truth-By-Consensus and \textsf{SVD}}
	\label{subsec:svd}
	
	\textit{Truth-By-Consensus} is Witnet's protocol for comparing and finding an "agreed truth" among a number of potentially conflicting claims brought by independent participants of the network.
	
	Truth-by-consensus is roughly the same algorithm as Truthcoin's\cite{truthcoin}, which in turn is based on \textit{Singular Value Decomposition} (\textsf{SVD}). This algorithm is in a way analogous to the statistical technique of Principal Component Analysis (PCA), although it introduces weighting in order to take reputation into account.
	
	The purpose of \textsf{SVD} is to analyze a matrix containing all the claims brought during one epoch and reveal and sort its effects by influence, detecting and dropping outliers and collusion (coordination) in the process.
	
	To measure coordination, \textsf{SVD} uses the first score from a weighted principal components analysis. This column represents the degree to which each miner's claims difer from those of a theoretical maximally representative of the covariance across all miners and claims.
	
	For each epoch, reputation points are redistributed among all the witnesses who were designated for fulfilling \textsf{RAD} tasks and brought their claims in a timely manner. As said before, if the claims are 100\% unanimous, reputation scores do not change (apart from the effect of demurrage).
	
	Reputation redistribution could become a rather expensive operation in terms of computational complexity as the number of network participants increase. To relieve this complexity, this protocol can be implemented in such a way that scores gets only updated once every few epochs. However, for security reasons, this "lazy reputation recomputation period" should be kept as low as possible as otherwise we would be deferring the punishment to liars and such measures would lose part of their efficacy. Please note that this recomputation period must form part of the network consensus as reputation score has important implications to the Useful Work Consensus algorithm in section~\ref{sec:pow}.
	
	With the view to ease the computational cost of reputation recomputation, the network shall keep a list of all participants whose reputation is different to the initial neutral value of $1$: the \textit{engaged set}. To limit the size of such set, all reputation scores with a value just above $1$ can be immediately assimilated to $1$, and the difference added to the reputation redistribution. As demurrage does not affect those miners with a neutral reputation, at any epoch checkpoint, only the reputation of the participants in the engaged set will need to be recomputed.
	
	Truth-by-consensus ensures that across a number of epochs, the network participants have a strong incentive to become witnesses, perform their assigned \textsf{RAD} tasks honestly and bring true claims: revenue maximization. All participants are incentivized to get and keep a high reputation as their potential income depends heavily on their score.
	
	As introduced earlier in section~\ref{sec:Witnet}, Witnet provides a strong incentive for witnesses to keep their claims secret until all of them have revealed their commitments, or even lie to each other or to a theoretical briber. This "double-agent incentive"---as named by Sztorc\cite{truthcoin}---guarantees that in the event that someone was trying to coordinate witnesses outside of the network, witnesses would prefer to lie to the coordinator and still perform their tasks honestly.
	
	As long as >50\% of the witnesses are honest, cartels and pools are heavily discouraged as each of the witnesses will want to minimize the number of fellow honest voters, as they all compete for a share of the same rewards and can thus be seen as rival.

	\newpage
	\section{Useful Work Consensus Algorithm}
	\label{sec:pow}
	
	The Proof-Of-Work (\textsf{PoW}) systems used by most public blockchains has been proved to be a great measure to achieve decentralized consensus and ultimately tell who gets the right to have the last word on which transactions are written into a common ledger.
	
	\textsf{PoW} puts a number of parties---\textit{the miners}---to compete for being the first to solve a mathematical problem whose solution is a "golden ticket" which grants its bearer the right to (1) aggregate as many transactions as they can fit into a limited-size block, (2) collect the unassigned value---\textit{the fees}---of those transactions, and (3) collect the block reward. Such mathematical problems are asymmetrical: they are extremely hard to solve, yet their solutions are easy to verify; and change every time a new valid block is published.
	
	While everyone can produce and broadcast a block, the network will only accept the first one that contains (1) a reference to last known valid block, and (2) valid a solution to the current problem. In short, your potential income as a miner depends on your computing power, which in turn depends on your ability to invest in specialized mining hardware and the price of electricity in your area.
	
	Given the nature of \textsf{DON}s, the chances of a certain miner to mine a block or to have a say in the outcome of a \textsf{RAD} request can not depend on its purchasing power or low price of electricity. Instead, those chances need to be related to reputation---past performance in terms of honesty---and, ultimately, randomness.
	
	The consensus scheme used by Witnet is similar to those proposed by Micali\cite{micali}, Bentov et.al.\cite{Bentov:2014:PAE:2695533.2695545}, Daian et.al.\cite{daian:2016:919}, and Benet et.al.\cite{filecoin}. In this scheme, the probability that the network elects a certain participant to create a new block or to fufill a \textsf{RAD} task is proportional to its reputation score in relation to the total reputation points existing in the network.
	
	\noindent\textbf{Miner influence}. In Witnet, the \textit{influence} $I^{t}_{i}$ of a participant $\mathcal{M}_i$ at checkpoint $t$ is the fraction of $\mathcal{M}_i$'s reputation score $r^t_i$ over the total reputation points in the network $\sum_{j}r^{t}_{j}$.
	
	\begin{figure}[H]
		\centering
		\[ I^{t}_{i} = \frac{r^t_i}{\sum_{j}r^{t}_{j}} \]
		\caption[Miner Influence Calculation]{\textbf{Miner influence calculation.} The influence of a participant in the network is proportional to its reputation stake. In that sense, reputation works as shares in a hypothetical "oracles corporation" formed by the set of all witnesses \cite{truthcoin}.}
		\label{fig:influence_calculation}
	\end{figure}
	
	One could directly substitute the $\sum_{j}r^{t}_{j}$ term in figure~\vref{fig:influence_calculation} with the total supply of reputation points. However, this would imply assuming that all the possible key pairs correspond to witnesses who are actively competing for leadership of the current epoch. Such assumption would cause new participants to have a ridiculously low chance to be elected for leadership ($\frac{1}{2^{256}}$ for \textsf{ECDSA}).
	
	Instead, the $\sum_{j}r^{t}_{j}$ term must be computed from the \textit{engaged set} as defined in section~\ref{subsec:svd}. This is, $r^{t}_{j} > 1$. This way, the influence of a certain participant is always proportional to its own reputation in comparison to the summation of that of its active peers.

	\subsection{Reputation-Based Mining Protocol}
	\label{subsec:mining}
	
	Witnet's Useful Work Consensus Algorithm aims to deterministically, unpredictably, and secretly elect a small set of miners at each epoch. Predictably, the number of elected miners per epoch is 1.
	
	This a probabilistic consensus protocol, where each epoch introduces more certainty over previous blocks, eventually reaching enough certainty that the likelihood of a different history is sufficiently small.
	
	A participant $\mathcal{M}_i$ is a miner at epoch $t$ if the following condition is met:
	
	\begin{figure}[H]
		\begin{center}
			\[ \mathcal{H}( \langle t \| \textsf{rand}(t) \rangle _{\mathcal{M}_i}) / 2^L \leq I^{t}_{i} \]
		\end{center}
		Where:
		\begin{itemize}
			\item \textsf{rand}($t$) is a public randomness that can be extracted from the blockchain at epoch $t$.
			\item $\langle t \| \textsf{rand}(t) \rangle _{\mathcal{M}_i}$ is a signature of message $t \| \textsf{rand}(t)$ produced with private key $_{\mathcal{M}_i}$.
			\item $\mathcal{H}$ is a deterministic, uniform and non-reversible hash function.
			\item $L$ is the number of bits of the output size of the $\mathcal{H}$ hash function.
			\item $I^{t}_{i}$ is the reputation of participant $\mathcal{M}_i$ at epoch $t$, calculated as in figure~\vref{fig:influence_calculation}.
		\end{itemize}
		\caption[Block Mining Calculation]{Block mining calculation.}
		\label{fig:mining_calculation}
	\end{figure}
	
	Being this protocol probabilistic, the number of miners per epoch can be expected to be 1, but only \textit{on average}. For some epochs there will be more than 1 miner, which is no problem at all for Witnet's ledger because of its \textsf{DAG} architecture. On the contrary, for some other epochs, it could happen that none of the participants would be eligible for mining\footnote{Or none of them realized their elegibility.}. Although empty blocks are possible (as described in section~\vref{subsubsec:ledger}), this protocol provides a \textit{"backup"} strategy that allows the network to avoid such empty blocks by letting a different subset of participants to take over the mining process if the former elected miners fail to produce and broadcast valid blocks.
	
	A participant $\mathcal{M}_i$ is a backup miner at epoch $t$ if the following condition is met:
	
	\begin{figure}[H]
		\begin{center}
			\[ \mathcal{H}( \langle t \| \textsf{rand}(t) \rangle _{\mathcal{M}_i}) / 2^L \leq \mathcal{B} \cdot I^{t}_{i} \]
		\end{center}
		Where:
		\begin{itemize}
			\item \textsf{rand}($t$), $\langle \dots \rangle _{\mathcal{M}_i}$, $\mathcal{H}$, $L$ and $I^{t}_{i}$ preserve the same meaning and value as in figure~\vref{fig:mining_calculation}.
			\item $\mathcal{B}$ is the backup index. Note that figure~\ref{fig:mining_calculation} is equivalent to this equation when $\mathcal{B} = 1$.
		\end{itemize}
		\caption[Backup Block Mining Calculation]{Backup block mining calculation.}
		\label{fig:backup_calculation}
	\end{figure}
	
	During each epoch, participants can claim their right to mine and produce blocks for any value of $\mathcal{B}$ such that $\mathcal{B} \in \mathbb{N}$ and $\mathcal{B} \geq 1$. Nevertheless, only those valid blocks with the lowest $\mathcal{B}$ value will be accepted and worked upon by the network. As producing a block is a trivial burden from a computational standpoint, it is to be expected that every participant will try to produce and broadcast valid blocks for different low values of $\mathcal{B}$, just in case no other participant succeeds to produce and broadcast a valid block for any lower value of $\mathcal{B}$.
	
	Methods for extracting randomness from public blockchains---as needed by \textsf{rand}($t$)---have already been proposed by Bonneau et.al.\cite{bonneau}. Note that the value of the randomness \textsf{rand}($t$) is derived from the hash of the blocks for epoch $t-1$ and therefore can not be known before epoch $t$.
	
	This type of scheme provides three main properties, as pointed out by Benet et.al.\cite{filecoin}:
	
	\begin{itemize}
		\item \textit{Fairness}: each participant has a fair chance to make a profit from their own work during each epoch, since signatures are deterministic and both $t$ and \textsf{rand}($t$) are fixed. Assuming $\mathcal{H}$ is a secure cryptographic hash function, then $\mathcal{H}( \langle t \| \textsf{rand}(t) \rangle _{\mathcal{M}_i}) / 2^L$ must be a real number uniformly chosen from the range [0, 1]. Hence, the probability for the equation to be true must be $I^{t}_{i}$, which equals to the participant's influence---its portion of reputation within the network. Because this probability is directly proportional to influence, this likelihood is preserved even under splitting or influence pooling.
		\item \textit{Secrecy}: an efficient adversary that does not own $\mathcal{M}_i$'s secret key can compute the signature with negligible probability, given the assumptions of digital signatures.
		\item \textit{Public verifiability:} a designated block miner can convince any wary verifier by showing $t$, \textsf{rand}($t$) and $\mathcal{H}(\langle t \| \textsf{rand}(t) \rangle _{\mathcal{M}_i})/2^{L}$. Given the previous point, no one can generate a proof without having a winning secret key.
	\end{itemize}
	
	\subsection{Reputation-Based Task Assignment Protocol}
	\label{subsec:assignment}
	
	The protocol that grants witnesses the right---and responsibility---to perform \textsf{Retrieve} and \textsf{Attest} tasks is quite similar to the Reputation-Based Mining Protocol described earlier in this section. Witnesses get higher or lower chances to be designated for fulfilling those tasks depending on their reputation score. Assignation of \textsf{Deliver} tasks to bridges is also ruled by this protocol and done by the same means.
	
	As introduced earlier in section~\ref{subsubsec:client_cycle}, every \textsf{RAD} request must contain a \textit{replication factor}, $\mathcal{R}$, that tells the network the minimum number of witnesses that must perform the \textsf{RAD} task and participate in the attestation.
	
	A witness $\mathcal{M}_i$ is elected for fulfilling a certain \textsf{RAD} task in epoch $t$ if the following condition is met:
	
	\begin{figure}[H]
		\begin{center}
			\[ \mathcal{H}( \langle t \| \textsf{rand}(t) \| n \rangle _{\mathcal{M}_i}) / 2^L \leq \mathcal{R} \cdot I^{t}_{i} \]
		\end{center}
		Where:
		\begin{itemize}
			\item $t$ is the epoch in which the \textit{time lock} of the request expires. If there is no time lock, the epoch in which the \textsf{RAD} request was mined into a block is used instead.
			\item \textsf{rand}($t$), $\langle \dots \rangle _{\mathcal{M}_i}$, $\mathcal{H}$, $L$ and $I^{t}_{i}$ preserve the same meaning and value as in figure~\vref{fig:mining_calculation}.
			\item $n$ is a flag that indicates the type of task (\textsf{RA} or \textsf{D}) and guarantees that no participant is assigned different types of tasks for the same request and epoch.
			\item $\mathcal{R}$ is the replication factor.
		\end{itemize}
		\caption[Task Assignment Calculation]{Task assignment calculation.}
		\label{fig:assignment_calculation}
	\end{figure}
	
	Being this a probabilistic protocol, even if a certain replication factor is specified, it can not guarantee that the number of actual witnesses getting assigned the task will match such factor. For example, for $\mathcal{R} = 5$ the number of witness could well be $0$ or $9$, but the thing is that most times---and in average---it will be $5$.
	
	If the number of witnesses who have discovered their assignment to a certain task is equal or greater than such task's replication factor, all of the witnesses are accepted.
	
	In the same manner, if the number of witnesses who discovered their assignment to a certain task is less than such task's replication factor, all the witnesses are accepted, but the "assignment contest" is not yet over. Instead, we must find the difference $\mathcal{D}$ between the replication factor and the actual number of designated witnesses and wait for next epoch $t + 1$. As soon as the new epoch begins, a witness $\mathcal{M}_i$ will be considered for "applying for the vacant positions" as soon as it can satisfy the condition from figure~\ref{fig:assignment_calculation} with $t = t + 1$ and $\mathcal{R} = \mathcal{D}$. This process will be repeated until the number of witnesses who have committed to the task is equal or greater than $\mathcal{R}$.
	
	\newpage
	\section{Transactions}
	\label{sec:transactions}
	
	\subsection{Transaction Properties}
	\label{subsec:tx_properties}
	
	\theoremstyle{definition}
	\begin{definition}{Transaction}
		: A Witnet transaction can be thought of as a pure function $f(s)$ such that, when applied on a valid ledger state $s_1$, results in a different valid ledger state $s_2$.
		\[ s_1 \neq s_2, s_1 \in{S} :  f(s_1) = s_2 \in S \]
	\end{definition}
	
	\theoremstyle{definition}
	\begin{definition}{Validity}
		: A transaction $f$ is valid over state $s_1$ when $s_1$ belongs to its domain.
		\[ f: s_1 \to s_2 \]
	\end{definition}
	
	\theoremstyle{definition}
	\begin{definition}{Independence}
		: Two transactions $f$ and $g$ are independent over state $s_1$ when both of them are valid over such state.
		\[ f: s_1 \to s_2 \text{ and } g: s_1 \to s_3 \]
	\end{definition}
	
	\theoremstyle{definition}
	\begin{definition}{Dependence}
		: Transaction $g$ depends on transaction $f$ over $s_1$ when it is not valid over $s_1$ but it is valid over $f(s_1)$. That is, the domain of $g$ is the codomain of $f$.
		\[ f: s_1 \to s_2 \text{ and } g: s_2 \to s_3 \]
	\end{definition}
	
	\theoremstyle{definition}
	\begin{definition}{Equivalence}
		: Two transactions $f$ and $g$ are equivalent over an initial state $s_1$ when the separate applications of each one of them over the initial state result in the same final state $s_2$.
		\[ f(s_1) = s_2 \text{ and } g(s_1) = s_2 \]
	\end{definition}
	
	\begin{theorem}[Completeness]
		For any initial state $s_1$ and final state $s_2$, being both different valid states, there exist one and only one transaction $f$ such that when applied on $s_1$ results on $s_2$.
		\[ s_1 \neq s_2, \forall{s_1} \in S, \forall{s_2} \in S : \exists!{f} \ni f(s_1) = s_2 \]
	\end{theorem}
	
	\begin{theorem}[Composition]
		For any finite set of transactions $F$, there exist one and only one transaction $g(s)$ that is equivalent to the composition of all the functions in the set.
		\[ \forall{F} = \{f_1, \cdot\cdot\cdot, f_{|F|}\} : \exists!g \ni (f_1 \circ \cdot\cdot\cdot \circ f_{|F|})(s) =  g(s) \]
	\end{theorem}
	
	\begin{theorem}
		For any initial state $s_1$ and final state $s_2$, being both valid states, there exist an infinite number of finite sets of transactions $F$ such that, when all their member transactions $f_n$ are composed and applied over $s_1$, result in $s_2$.
		\[ s_1 \neq s_2, \forall{s_1}\in{S}, \forall{s_2}\in{S} : \exists^{\infty}F \ni {\underset {n=1}{\overset {|F|}{\mathop {R}}}}\,f_n(s_1) = (f_1 \circ \cdot\cdot\cdot \circ f_{|F|})(s_1) = s_2 \]
	\end{theorem}
	
	\begin{theorem}[Commutativity]
		For any set of transactions $F$, composition of all its member transactions $f_n \in F$ is a commutative operation if and only if all of its members are independent to each other. 
		\[ \forall{F} = \{f_1, \cdot\cdot\cdot, f_{|F|}\} : (f_1 \circ \cdot\cdot\cdot \circ f_{|F|})(s) = (f_{|F|} \circ \cdot\cdot\cdot \circ g_1)(s) \]
		Put more simply: \[ f_1(f_2(s)) = f_2(f_1(s)) \]
	\end{theorem}
	
	\subsection{Types of Transactions and Standard Outputs}
	\label{subsec:tx_types}
	
	There exist at least 4 types of "standard" transactions:
	
	\begin{itemize}
		\item \textbf{Value transfer transactions} (\textsf{VTT}). Roughly equivalent to Bitcoin's \textsf{P2PKH} and \textsf{P2SH}.
		\item \textbf{\textsf{Client RAD request} transactions}. Codified \textsf{RAD} requests. They contain the retrieval paths, the aggregator function and optionally one or many \textsf{deliver} clauses.
		\item \textbf{Witness \textsf{commit-pledge} transactions}. Used by witnesses to (1) commit the results of their retrieval tasks without revealing the actual claims, and (2) pledge their portion of the witness reward.
		\item \textbf{Witness \textsf{reveal-redeem} transactions}. Used by witnesses to (1) reveal the actual claims that they committed in their \textsf{commit-pledge} transactions, and (2) redeem their portion of the witness reward.
	\end{itemize}
	
	When sent over the network, all of these types of transactions are encoded using the same serialization format. Also, all the types of transactions use the same output scripting language described in section~\vref{subsec:tx_outputs}.
	
	\subsection{Fees}
	\label{subsec:fees}
	
	\subsubsection{Miner fees}
	\label{subsubsec:miner_fees}
	
	Miner fees work just like in other public blockchain protocols: being block space a very limited resource, transactors need to compete for it. Thus, their only means for persuading block miners to include their transactions before others' is to pay a miner fee higher than the rest's.
	
	As seen with Bitcoin, it is to be expected that as long as the block size limit is not reached, the miner fees will be kept low. Then, as blocks start to be full, a transaction backlog is formed and fees start to raise so that superflous (or even spam-\textit{ish}) transactions are disincentivized and no longer made, which translates to more available disk space and eventually lower fees. Over time, this cycle ends up striking a balance in which (1) blocks are full, and (2) miner fees only grow at the same pace that demand for block space does.
	
	Since in  Witnet's case miners do not need to perform any extremely expensive operation to mine blocks, they are not urged to arbitrarily reject low-fee or even zero-fee transactions. As mentioned earlier, it is obvious that they will always try to prioritize best-paying transactions. However, as long as all the pending transactions fit in a single block, there is no reason for a block miner to not include them all.
	
	This perfectly matches Satoshi's vision on miner fees: \textit{The fee the market would settle on should be minimal. If a node requires a higher fee, that node would be passing up all transactions with lower fees. It could do more volume and probably make more money by processing as many paying transactions as it can. The transition is not controlled by some human in charge of the system though, just individuals reacting on their own to market forces}\cite{satoshi:fees}.
	
	It is worth remembering that mining is initially subsidized by the \textsf{Wit} generation algorithm as described in section~\ref{subsubsec:wit}. This is, block miners do not only obtain tokens from the miner fees attached to the transactions they process, but also get a fixed amount for every block they mine---the block reward. As the inflation will decay over time due to block reward being halved periodically, mining will be less subsidized on the long run and miner fees will gain importance.
	
	\subsubsection{Witness fees}
	\label{subsubsec:witness_fees}
	
	As mentioned earlier in this work, witness fees calculation is a function of:
	
	\begin{itemize}
		\item Number of \textbf{acquisition paths}, as defined in section~\ref{subsubsec:crosscheck}.
		\item Upper bound of \textbf{computational complexity} of each of the normalization scripts in the acquisition paths.
		\item The \textbf{replication factor}, $\mathcal{R}$, required by the client.
	\end{itemize}
	
	This calculation tells the client how expensive will a \textsf{RAD} request for the network to fulfill. As the computational power of the network for every epoch is rather limited, this function offers an absolute metric that is similar to that of transaction size in comparison to block size.
	
	\begin{figure}[H]
		\centering
		\[ fee \propto \mathcal{R} \cdot \sum_{n = 0}^{|paths|} \mathcal{O}(paths_{n}) \]
		\caption[Witness Fees Calculation]{Witness fees are proportional to the required replication factor and the total computational complexity of the acquisition paths ($\mathcal{O(\dots)}$).}
		\label{fig:witness_fees}
	\end{figure}
	
	The actual cost for witnesses to perform the tasks they get assigned is marginal and virtually negligible when compared to the rewards they get in exchange. The only significant costs involved will be the miner fees they will need to pay to broadcast their \textsf{commit-pledge} plus \textsf{reveal-redeem} transactions and eventually collect their rewards.
	
	For each \textsf{RAD} request, each witness' reward will always equal the request's witness fee divided by the number of witnesses who commit their solutions to the request, which in turn shall be equal or greater than the replication factor.
	
	For these reasons, as long as (1) a witness has spare computational power and bandwidth, and (2) the witness reward is above a certain threshold---greater than twice the miner fees---, there is no reason why a witness node would opt to ignore a single task it got designated for, all the more since such laziness is heavily penalized by the reputation demurrage mechanism introduced in section~\ref{subsec:reputation_protocol}.
	
	It is also worth stating that Witnet's reputation scheme somehow turns miner fees into bonuses for which only honest witnesses are eligible. This means that the profit that witnesses receive in exchange for their work does not only come directly from the witness fees that clients pay, but also indirectly from the earned reputation points, which eventually give them a higher chance to mine a block and collect block rewards, including miner fees. Theoretically, this fact could even lead some witnesses to accept tasks at a loss in the expectation that long term block rewards could compensate and exceed the loss.
	
	\subsubsection{Bridge fees}
	\label{subsubsec:bridge_fees}
	
	Bridge fees follow the same criteria as witness fees. 
	
	\subsection{Outputs and Scripting}
	\label{subsec:tx_outputs}
	
	Witnet is conceived to use a scripting language to lock tokens in outputs, akin to Bitcoin's \textsf{script} language\cite{bitcoin:script}. This language--called \textsf{WitScript}--is used to specify under which conditions the tokens locked in a transaction output can be redeemed, and in most cases, by whom. These scripts embedded in transactions is what we commonly call \textit{smart contracts}.
	
	In the light of the successful activation of segregated witnesses (\textit{segwit}) in the Bitcoin protocol and the countless benefits it supposed for the health of the network, we consider that such set of features should be available in the Witnet protocol from day one.
	
	Adopting \textit{segwit} from the very beggining opens the door to supporting multiple smart contract languages that go beyond \textsf{WitScript}, like EVM\cite{ethereum}, Michelson\cite{michelson} or Simplicity\cite{simplicity}. While these languages are still low-level, they are closer to be general-purpose and can be used as compilation targets for popular high-level smart contract languages, such as Solidity\cite{solidity} or Ivy\cite{ivy}.
	
	Using different \textit{segwit} version numbers also leaves space for a variety of signature schemes that can offer greater flexibility, scalability and privacy than \textsf{ECDSA}, as is the case with Schnorr signatures\cite{schnorr}.

	\subsubsection{MAST and Tail Call Execution}
	\label{subsubsec:tail_call}
	
	To provide smart contracts with greater flexibility, \textsf{WitScript} shall support \textit{Merklized Abstract Syntax Trees} (MAST). This feature, as proposed by Robin et.al\cite{robin:mast}, Lau\cite{lau:mast,lau:mastopcodes} and Friedenbach et.al.\cite{friedenbach:fastmerkle,friedenbach:merklebranchverify,friedenbach:tailcall} enables decomposition of complex branching scripts into a set of non-branching flat execution pathways, committing to the entire set of possible pathways, and then revealing only the path used at spend time.
	
	One fundamental part of MAST is \textit{Tail Call Execution}. It changes the behaviour of the output scripts interpreter in such a way that if the memory stack is not clean at the end of script execution, the remaining elements in the stack will (1) be treated as serialized scripts and inputs, (2) immediately executed in a secondary stack, and (3) finally replaced in the main stack by the result of the execution. Different types of \textit{tail scripts} can exist to implement different behaviors.
	
	\subsubsection{Covenants}
	\label{subsubsec:covenants}
	
	To enable and assure the \textsf{commit-pledge} and \textsf{reveal-redeem} transactions scheme described earlier in section~\vref{subsubsec:miner_cycle}, \textsf{WitScript} implements a special type of tail script that, when used to lock an output, forces spending transactions to have only outputs that abide by a certain "template". This feature, called \textit{covenants}\cite{moser:covenants}\cite{oconnor:covenants}, is a kind of \textit{"forced script heirship"}: inputs can cause outputs to inherit a certain policy---in form of script---that the spender can not help but honor.
	
	Covenant clauses are copied in front of all the transaction outputs' redeem scripts in the same order as they appear in the inputs. A transaction spending outputs affected by covenants but missing to pass such clauses to its outputs is considered not valid.
	
	Thanks to this feature, \textsf{RAD} requests can require spending transactions to be \textsf{commit-pledge} transactions that can only be spent by \textsf{reveal-redeem} transactions.
	
	While incredibly useful for many use cases, covenants must be carefully implemented by user-facing client applications such as wallets. These applications must double-check that incoming value transactions are free from covenant clauses before showing them as valid value inputs and updating the user's balance accordingly. Otherwise, a malicious payer could induce a gullible recipient to believe that the value of a certain payment is freely spendable, while in fact it can only be spent in a certain way predefined by the malicious payer.
	
	\subsubsection{Bundled Macros}
	\label{subsubsec:tail_call}
	
	\textsf{RAD} requests, \textsf{commit-pledge} and \textsf{reveal-redeem} transactions are significantly bigger in bytesize that regular value transfer transactions. In anticipation of a significant part of the limited block space being taken by those types of transactions, \textsf{WitScript} provides a series of macros that implement the most frequent transaction formats so that they take less space when stored or sent over the network.
	
	Every macro is equivalent to a predefined \textsf{WitScript} code. For illustrative purposes, figure~\vref{example:op_request} depicts usage of a theoretical \textsf{M\_REQUEST} macro and its equivalent \textsf{WitScript} program. Please note that this is just an example and the final \textsf{M\_REQUEST} macro used in a first implementation of Witnet will not necessarily need to be exactly like this.
	
	\begin{figure}[H]
		\begin{example}[example:op_request]
			<$client\_key\_hash$> <$replication\_factor$> <$rad\_request\_bin$> M\_REQUEST
			\tcblower
			<$rad\_request\_bin$> OP\_DEPLOY\\
			OP\_DUP OP\_HASH160 <$client\_publicKeyHash$> OP\_EQUAL\\
			OP\_IF\\
			\hspace*{2em} 10 OP\_CHECKLOCKTIMEVERIFY
			OP\_ELSE\\
			\hspace*{2em} OP\_DUP <$replication\_factor$> OP\_CHECKMINERVERIFY\\
			\hspace*{2em} <$M\_PLEDGE$> <$COVENANT\_FLAG$> \# These are serialized for tail call execution\\
			OP\_ENDIF\\
			OP\_DROP OP\_CHECKSIGVERIFY
		\end{example}
		\caption[\textsf{M\_REQUEST} Macro]{\textbf{The \textsf{M\_REQUEST} macro}. It allows elected miners to pledge a fraction of the locked funds as long as all outputs in the pledging transaction start with the \textsf{M\_PLEDGE} macro. If not pledged after 10 blocks, the original requester can apply for a refund.}
		\label{fig:op_request}
	\end{figure}
	
	When the \textsf{WitScript} interpreter bumps into a macro, it will not just substitute it by its equivalent \textsf{WitScript} code. Instead, the interpreter will:
	
	\begin{enumerate}
		\item Pop as many elements from the stack as the number of parameters required by the macro.
		\item Run a precompiled function that implements the very same logic than the macro being executed, passing the popped elements as arguments to such function.
		\item Push the return value of the function to the stack.
		\item Keep interpreting the \textsf{WitScript}.
	\end{enumerate}
	
	This macro optimization functionality is inspired in the concept of \textit{"jets"} as introduced by O'Connor\cite{simplicity} and Yarvin et.al\cite{urbit}. Macros not only save block space and bandwidth, but also can make \textsf{WitScript} run faster, as many resource-intensive parts such as signature verification can be delegated to precompiled, optimized and formally verified modules.
	
	\newpage
	\section{Bridges and Smart Contracts}
	\label{sec:sc}
	
	\subsection{Bridges: Interacting with other Platforms}
	\label{subsec:bridges}
	
	Bridge nodes, as introduced before in section~\vref{fig:participants}, aim to connect Witnet to other platforms. Of special note is the connection of the Witnet network to other public smart contract platforms like Ethereum\cite{ethereum}, Decentralized Storage Networks such as Filecoin\cite{filecoin} and directly to the web.
	
	Bridges targetting different platforms also use different values for the $n$ parameter in the Reputation-Based Task Assignment Protocol function as defined by figure~\vref{fig:assignment_calculation}.
	
	\subsubsection{Ethereum Bridges}
	\label{subsubsec:ethereum_bridges}
	
	Ethereum bridges are Witnet bridge nodes which also run an Ethereum full node, have full access to the Ethereum blockchain and have the capability to operate with \textsf{ether} and make contract calls. Ethereum bridges are in charge of two missions:
	
	\begin{itemize}
		\item \textbf{Requests introduction}. Ethereum bridges monitor the Ethereum blockchain in search for \textsf{RAD} requests codified inside regular Ethereum transactions. When they find one of these transactions, they read the payload and convert it into a valid Witnet \textsf{RAD} request that they must broadcast to the Witnet network. In this scheme, the bridge will act as an intermediate client between the actual client (who is an unknown Ethereum account or contract) and Witnet. In exchange of performing this work and spending their own \textsf{Wit}s, bridges are rewarded by the Ethereum clients using \textsf{ether} (or any other Ethereum token).
		
		\item \textbf{Results reporting}. Ethereum bridges are also in charge of reporting the results of those \textsf{RAD} requests which specify Ethereum-targetted \textsf{Delivery} clauses to the Ethereum blockchain. In exchange of performing this work and spending their own \textsf{ether} by attaching gas to their report transactions, they are rewarded with \textsf{Wit} tokens that were allocated for such purpose in the request transaction.
	\end{itemize}
	
	Ethereum clients will need to make sure they attach enough value to their requests as to reward not only the bridge who will introduce the request into Witnet but also the one who will report the result back the client.
	
	Ethereum bridges are expected to fulfill an important role for adoption of Witnet as they will ease the integration with the thriving Ethereum smart contract ecosystem. However, the economics behind this scheme are relatively fragile as significant fluctuations between the exchange rate of \textsf{Wit} and \textsf{ether} tokens may render it unsuitable for long-lived contracts. For this reason---in addition to cost-effectiveness and lower latency---the preferred way to programatically transfer value depending on the result of a \textsf{RAD} request will be using Witnet native smart contracts, as explained in section~\vref{subsec:native_sc}.
	
	\subsubsection{Decentralized Storage Networks (\textsf{DSN}) Bridges}
	\label{subsubsec:dsn_bridges}
	
	Decentralized Storage Networks (\textsf{DSN}) aggregate storage offered by multiple independent storage providers and self-coordinate to provide data storage and data retrieval to clients\cite{filecoin}. 
	
	\textsf{DSN} bridges provide means for (1) storing the results of \textsf{RAD} requests into \textsf{DSN}s, as well as (2) ensure that those results are persisted in those \textsf{DSN}s in perpetuity.
	
	This scheme and one of its most interesting use cases---ensuring that access to truth and human knowledge will remain democratic forever---are further explained in appendix~\vref{sec:ark}.
	
	\subsubsection{Web Bridges}
	\label{subsubsec:web_bridges}
	
	Just like other types of bridges, they act as intermediaries who receive \textsf{RAD} requests from third parties and post them into Witnet. 
	
	Web bridges are specially convenient for integrating Witnet with traditional web technologies, as they provide:
	
	\begin{itemize}
		\item A public endpoint that third parties use to send their requests.
		\item A public endpoint that third parties use to read the result of their requests.
		\item A WebHook\footnote{Webhooks are "user-defined HTTP callbacks". They are usually triggered by some event, such as pushing code to a repository or a comment being posted to a blog. When that event occurs, the source site makes an HTTP request to the URL configured for the webhook. --- Wikipedia.} or EventSource\cite{eventsource} service that notifies clients when the results of their requests are ready.
	\end{itemize}
	
	These two components can be either APIs, user interfaces or both. Note that they are quite similar to block explorers, and it is to be expected that many web bridges will also act as block explorers and vice versa.
	
	Participants running web bridges are free to charge their users for their services in whatever form they want.
	
	\subsection{Native Smart Contracts}
	\label{subsec:native_sc}
	
	The core functionality of Witnet as a \textsf{DON} provides an efficient way to retrieve, attest and deliver verified web contents. This alone is enough to enable the creation of countless novel applications capable of reacting to the outer world without human intervention or relying on single, centralized---thus corruptible or hackable---sources of information.
	
	However, an even broader spectrum of use cases and decentralized applications are made possible if the \textsf{DON} provides a smart contract language so that \textsf{Wit} tokens themselves are programmable and can react to the results of \textsf{RAD} requests.
	
	Witnet implements a smart contract feature that inherits its approach from Bitcoin's Script. Contracts in Witnet are stateless programs that regulate when, how and by whom certain funds can be spent.
	
	Any \textsf{WitScript} used to lock a transaction output can read the result of a certain \textsf{RAD} request and push it into the execution stack by using the \textsf{OP\_READ} operation code.
	
	The example~\vref{example:smart_contract} depicts a smart contract that, in spite of being really simple, was impossible to implement in a trustless way until now.
	
	\begin{figure}[H]
		\begin{example}[example:smart_contract]
			A transaction output is set to be spendable by one of two different parties, depending on a certain \textsf{RAD request} returning a result value above or under 50,000.
			\tcblower
			OP\_HASH160 <$RAD\_request\_id$> OP\_READ 50000 OP\_GREATERTHANOREQUAL
			OP\_IF\\
			\hspace*{2em} <$participant_{a}\_pubKeyHash$>\\
			OP\_ELSE\\
			\hspace*{2em} <$participant_{b}\_pubKeyHash$>\\
			OP\_ENDIF\\
			OP\_EQUALVERIFY OP\_CHECKSIG
		\end{example}
		\caption[Witnet Smart Contract Example]{Witnet smart contract example.}
		\label{fig:smart_contract}
	\end{figure}

	\newpage
	\begin{appendices}
		\section{The Digital Knowledge Ark}
		\label{sec:ark}
		\renewcommand{\thesubsection}{\Alph{section}.\alph{subsection}}
		
		\subsection{Motivation}
		\label{subsec:ark_motivation}
		
		\begin{center}
			\textit{\textbf{"History is written by the victors."}}
			\\Winston Churchill
		\end{center}
		
		\lettrine[nindent=0.5em,lines=2]{C}{hurchill's} famous dictum may appear to no longer hold true in an age where the Internet wields enormous potential for any person to share their beliefs and opinions with the rest of the world. However, centralized systems for the archiving of human knowledge are still very vulnerable to manipulation or destruction by corrupt governments and other malicious actors who could greatly benefit from altering history.
		
		As a society we have the responsibility to find a better way to preserve our cultural heritage from any odds that the future may hold. And we need it to be highly resilient, decentralized, self-governed and censorship-resistant, guaranteeing that knowledge will be accessible to everyone, everywhere, at any time, without discrimination of any type. Only in this way we will be able to ensure that access to human knowledge will remain democratic forever.
		
		As explained in section~\vref{subsubsec:verifiability}, verifiability is a weaker notion than truth. In this Information Age, verifiability is indeed very fragile because of the ephemerality inherent to digital media: one statement can be verifiable right now but lose its verifiability right after.
		
		\begin{example}[example:wikipedia]
			Let us visualize how fragile verifiability is:
			\begin{enumerate}
				\item Open a random Wikipedia article.
				\item Scroll all the way down to the bottom of the page.
				\item Follow all the links in the \textit{References} and \textit{External links} sections.
			\end{enumerate}
			\tcblower
			The chances are that at least one of the links is broken or does not point to the actual content it was supposed to.
		\end{example}
		
		In a similar way to Witnet\footnote{See subsection~\ref{subsubsec:verifiability}}, Wikipedia has a strong policy stating that the only valid truth is the one you can verify\cite{wiki:verifiability}. Nevertheless, due to the ephemeral nature of the web, even a well documented Wikipedia article may be rendered questionable or even not suitable under the Wikipedia policies and guidelines if many of its sources disappeared from the web.
		
		We are clearly in need of some infallible way to ensure that something that is verifiable today will still be verifiable tomorrow.
		
		The current most popular approach to preserving the availability of some data is using the Wayback Machine service run by the Internet Archive initiative to request an on-demand snapshot of any web site in which the data is published. However, centralized solutions offer little to no guarantee of the very ingredients that make contents verifiable:
		
		\begin{itemize}
			\item Content integrity: equivalence between the content in the snapshot and the actual content published on the web at the time it was taken\footnote{Assuming that the trusted attesting third party will not misbehave, content integrity will still be broken if attestations are made while the systems or networks of the trusted attesting party are under control of an attacker (malware, DNS spoofing/poisoning, broken TLS encryption, etc). Only viable solution to this problem is performing the attestations by using several "witnesses" in a decentralized way, just like the Witnet miners do.}.
			\item Custody integrity: equivalence between the original content in the snapshot and the one presented when retrieving such snapshot after some time\footnote{Assuming that the trusted attesting third party will not misbehave, custody integrity can still be broken if the systems or networks of the trusted attesting party are under control of an attacker at any time after the attestation. Only viable solution to this problem is persisting the attested contents in Decentralized Storage Networks (\textsf{DSN}) based on public blockchains, which is the approach of the Digital Knowledge Ark we are proposing here.}.
		\end{itemize}
		
		We would like to emphasize that our point here is not calling anyone's honesty into doubt. Our point is to stress the fact that such a single source of truth, no matter how reliable it is, also represents a single point of failure that introduces the chance for external malicious actors to rewrite or delete part of the history by breaking into a single system or network.
		
		\subsection{Knowledge Commons and Commons-based Peer Production}
		\label{subsec:commons}
		
		The commons is the cultural and natural resources accessible to all members of a society and held in common, not owned privately. Although the term originally referred to common land, nowadays it is taken to mean any shared and unregulated resource such as atmosphere, oceans, rivers, fish stocks, or even an office refrigerator.
		
		Those resources identified as commons are often said to be vulnerable to social dilemmas\cite{huberman} and governance problems that lead to competition for use, free riding, commodification, pollution, degradation, and ultimately non-sustainability\cite{hess}. These dilemmas are highlighted by the \textit{the tragedy of the commons}.
		
		The concept of \textit{tragedy of the commons} was introduced by ecologist Garrett Hardin in a 1968 article of the same name\cite{hardin}, inspired itself by an essay written in 1833 by the Victorian economist William Forster Lloyd\cite{lloyd}.
		
		The tragedy illustrates the argument that free access to a finite resource ultimately damage the resource through over-exploitation, temporarily or permanently. This occurs because the benefits of exploitation make some of the users want to maximize their own use while the costs of the exploitation are borne by everyone. This, in turn, causes demand for the resource to increase, which causes the problem to snowball until the resource collapses.
		
		However, knowledge forms part of a different kind of commons: non-substractible ones. Unlike environmental commons---e.g. a water spring---multiple users can access the same resource with no negative effect on its quality or quantity. When a teacher gives a lesson, knowledge is not split between the students but replicated across the minds of all of them.
		
		While substractible and non-substractible commons are similar in their shared nature, there is a radical difference in the source of their value as resources. The value of substractible resources is based on scarcity, whilst the value of non-substractible resources is based on abundance.
		
		Likewise, \textit{preservation} of substractible and non-substractible commons involve very different actions. Preservation of a substractible commons often means guaranteeing its availability by regulating access or imposing\footnote{In many cases, \textit{self-imposing}.} reasonable use rules on the resources, effectively making it somewhat less open to the public. On the contrary, preservation of a non-substractible commons means guaranteeing its availability by making it accessible to the greatest number of people and effectively making it more open to the public. Preservation is further discussed in section~\vref{subsec:preservation}.
		
		The culture heritage mentioned in section~\vref{subsec:ark_motivation} is none other than the \textit{knowledge commons}: the set of all knowledge and wisdom that our civilization has accumulated over the centuries and belongs to the whole of humanity. The knowledge commons is our legacy from the past, what we know today, and what we will pass on to future generations.
		
		The proposed Digital Knowledge Ark aims to form part itself of the \textit{digital commons} as defined by social researcher Mayo Fuster: \textit{"information and knowledge resources that are collectively created and owned or shared between or among a community and that tend to be non-exclusive, that is, be (generally freely) available to third parties. Thus, they are oriented to favor use and reuse, rather than to exchange as a commodity. Additionally, the community of people building them can intervene in the governing of their interaction processes and of their shared resources"}\cite{fuster}.
		
		The proposed Digital Knowledge Ark intends to leverage the current profusion of emerging blockchain technologies and projects to engage the people in the enrichment and preservation of the knowledge commons. It is conceived as a commons-based peer production initiative as defined by professor Yochai Benkler: \textit{"collaboration among large groups of individuals [...] who cooperate effectively to provide information, knowledge or cultural goods without relying on either market pricing or managerial hierarchies to coordinate their common enterprise."}\cite{benkler}.
		
		\subsection{Preservation and its Principles}
		\label{subsec:preservation}
		
		As mentioned earlier in section~\ref{subsec:commons}, preservation of knowledge commons entails guaranteeing the availability of knowledge resources by making them accessible to the greatest number of people, effectively making those resources more open to the public.
		
		There is one program from the Stanford University Libraries that we would like to acknowledge here for its notable contribution to defining the principles of long-term preservation of knowledge commons. Its name is pretty explicit about what is the cornerstone of its vision: "\textit{Lots Of Copies Keep Stuff Safe} (LOCKSS)"\cite{lockss:about}.
		
		The preservation principles of the LOCKSS program\cite{lockss:principles}---which we endorse and make them our own---focus on:
		\begin{itemize}
			\item Decentralized and distributed preservation.
			\item Preservation of original content.
			\item Perpetual, guaranteed and seamless access.
			\item Affordability and sustainability.
		\end{itemize}
		
		The Digital Knowledge Ark proposed here pays special attention to those very same principles, making the most of Decentralized Oracle Networks (\textsf{DON}), Decentralized Storage Networks (\textsf{DSN}) and other blockchain technologies to ensure their effective attainment.
		
		\subsection{Using Witnet to Agree on Facts to be Preserved}
		\label{subsec:Witnet_for_ark}
		
		Anyone interested in storing information in the Digital Knowledge Ark shall:
		
		\begin{enumerate}
			\item Create a \textsf{RAD} request with one or more valid retrieval paths pointing to the source of such information.
			\item Add a \textsf{deliver} clause to the request. This clause will ask for publication in a \textsf{DSN}.
			\item Fund the transaction with an amount of \textsf{Wit} tokens enough to reward miners, witnesses and bridges. Fees have been discussed in section~\vref{subsec:fees}.
			\item Send the \textsf{RAD} request to the network, either directly as a client or through bridge nodes.
		\end{enumerate}
		
		Provided that all these points are met, the information will be retrieved and attested by the Witnet \textsf{DON}, and bridge nodes will publish it into the \textsf{DSN} of choice.
		
		As with bridges targetting different platforms, \textsf{DSN} bridge nodes use different values for the $n$ parameter in the Reputation-Based Task Assignment Protocol function as defined by figure~\vref{fig:assignment_calculation}.

		\subsection{Persisting Facts and Contents in Perpetuity}
		\label{subsec:perpetuity}
		
		\lettrine[nindent=0.5em,lines=2]{P}{erpetual} storage is possible in one way or another in most existing public blockchains.
		
		For example, Ethereum smart contract can be used to keep data inside their state, which can be simply organized in a key-value mapping that allows entering a new record by calling an external function. The contract must calculate the hash of the data, use the hash as the key for storing the data and return the hash to the sender. Then the sender or anyone else who knows the hash can retrieve the data by calling a constant function using the hash as its sole parameter. Note that although data retrieval is made through a constant function and therefore can be performed an unlimited number of times at zero gas cost for the requesting party, data storage has side-effects (modifies state). This implies that the cost of the storage request increases linearly with the size of the data to be stored. While it can be a very acceptable solution for storing small data units in perpetuity, at scale, storing bigger data units (even in the order of a few kilobytes) becomes really expensive and impractical in most cases.
		
		Instead, archiving of big data units should be done by using decentralized storage solutions specially designed for storage of high volumes of data.
		
		There are already a bunch of promising projects that could be used for that purpose. Among the ones that better fit the requirements of the Ark, five deserve special mention here:
		
		\begin{itemize}
			\item InterPlanetary File System (IPFS)\cite{ipfs}: a peer-to-peer distributed file system that seeks to connect all computing devices with the same system of files.
			\item FileCoin\cite{filecoin}: a distributed electronic currency whose nodes are incentivized to store as much of the entire network’s data as they can, using the IPFS protocol.
			\item Sia\cite{sia}: a platform for decentralized storage in which peers can freely form blockchain-based storage contracts in a free and open market.
			\item Storj\cite{storj}: a peer-to-peer cloud storage network implementing client-side encryption without reliance on a third party storage provider.
			\item Swarm\cite{swarm1}\cite{swarm2}: a decentralized and redundant store of Ethereum’s public record, in particular to store and distribute dapp code and data as well as block chain data.
		\end{itemize}
		
		Interaction between the Witnet blockchain and these other networks will be made possible by \textsf{DSN} bridge nodes as introduced in section~\vref{subsubsec:dsn_bridges}.
		
		None of the aforementioned systems offer perpetual storage per se. Instead, they allow for establishing the duration of the storage contract. The more is paid, the longer the data will be persisted. The cost for storing a certain amount of data for a defined period of time is driven by supply and demand, and different nodes compete on factors like reliability and price.
		
		Ensuring that a certain data unit is never deleted from the decentralized storage system of choice thus imply recurrent costs. In order to impede deletion of data included in the Ark, interested clients and \textsf{DSN} bridges shall maintain an index that will relate all the archived data to the address of their corresponding storage contract and its date of expiry.
		
		The same clients that originally requested the retrieval of the information and the formalization of a storage contract can keep sending additional tokens to the storage contract to keep it in force. In addition, shall those indexes be publicly available, any other interested party could extend the storage contract by independently funding it.
		
		In essence, as long as there are enough actors interested in maintaining humanity’s most relevant digital data preserved in the Ark, we can be certain that access to our cultural heritage and legacy will remain democratic forever.
	\end{appendices}
	
	\clearpage
	\bibliographystyle{ieeetr}
	\bibliography{main}

\begin{thebibliography}{10}

\bibitem{augur}
J.~Peterson and J.~Krug, ``Augur: a decentralized, open-source platform for
  prediction markets,'' {\em CoRR}, vol.~abs/1501.01042, 2015.
\newblock \url{http://arxiv.org/abs/1501.01042}.

\bibitem{gnosis}
M.~{Köppelmann et. al.}, ``Gnosis: Crowdsourced wisdom,'' 2017.
\newblock \url{https://gnosis.pm/resources/default/pdf/gnosis_whitepaper.pdf}.

\bibitem{delphi}
Anonymous, ``Delphi,'' 2017.
\newblock \url{https://delphi.markets/whitepaper.pdf}.

\bibitem{bitcoin:paper}
S.~Nakamoto, ``Bitcoin: A peer-to-peer electronic cash system,'' 2009.
\newblock \url{https://bitcoin.org/bitcoin.pdf}.

\bibitem{truthcoin}
P.~Sztorc, ``Truthcoin: Peer-to-peer oracle system and prediction
  marketplace,'' 2015.
\newblock \url{http://www.truthcoin.info/papers/truthcoin-whitepaper.pdf}.

\bibitem{schellingcoin}
V.~Buterin, ``Schellingcoin: A minimal trust universal data feed,'' 2014.
\newblock
  \url{https://blog.ethereum.org/2014/03/28/schellingcoin-a-minimal-trust-universal-data-feed/}.

\bibitem{schelling}
T.~Schelling, {\em The Strategy of Conflict}.
\newblock Harvard University Press, 1960.

\bibitem{bendahmane}
A.~Bendahmane, M.~Essaaidi, A.~E. Moussaoui, and A.~Younes, ``The effectiveness
  of reputation-based voting for collusion tolerance in large-scale grids,''
  {\em IEEE Transactions on Dependable and Secure Computing}, vol.~12,
  pp.~665--674, Nov 2015.
\newblock \url{http://ieeexplore.ieee.org/document/6951404/}.

\bibitem{watanabe}
K.~Watanabe, N.~Funabiki, T.~Nakanishi, and M.~Fukushi, ``Modeling and
  performance evaluation of colluding attack in volunteer computing systems,''
  in {\em Proceedings of the International MultiConference of Engineers and
  Computer Scientists}, vol.~II, 2012.
\newblock
  \url{http://www.iaeng.org/publication/IMECS2012/IMECS2012_pp1658-1663.pdf}.

\bibitem{damiani}
E.~Damiani, D.~C. di~Vimercati, S.~Paraboschi, P.~Samarati, and F.~Violante,
  ``A reputation-based approach for choosing reliable resources in peer-to-peer
  networks,'' in {\em Proceedings of the 9th ACM Conference on Computer and
  Communications Security}, CCS '02, (New York, NY, USA), pp.~207--216, ACM,
  2002.
\newblock \url{http://doi.acm.org/10.1145/586110.586138}.

\bibitem{xiong}
L.~Xiong and L.~Liu, ``Peertrust: supporting reputation-based trust for
  peer-to-peer electronic communities,'' {\em IEEE Transactions on Knowledge
  and Data Engineering}, vol.~16, pp.~843--857, July 2004.
\newblock \url{http://ieeexplore.ieee.org/abstract/document/1318566/}.

\bibitem{porter}
R.~H. Porter, ``Detecting collusion,'' {\em Review of Industrial Organization},
  vol.~26, no.~2, pp.~147--167, 2005.
\newblock \url{http://www.jstor.org/stable/41799228}.

\bibitem{lepinski1}
M.~Lepinksi, S.~Micali, and A.~Shelat, ``Collusion-free protocols,'' in {\em
  Proceedings of the Thirty-seventh Annual ACM Symposium on Theory of
  Computing}, STOC '05, (New York, NY, USA), pp.~543--552, ACM, 2005.
\newblock \url{http://doi.acm.org/10.1145/1060590.1060671}.

\bibitem{lepinski2}
M.~Lepinski, {\em Steganography and collusion in cryptographic protocols}.
\newblock PhD thesis, Massachusetts Institute of Technology, 2006.
\newblock \url{http://hdl.handle.net/1721.1/38309}.

\bibitem{shareef}
A.~Shareef, ``Collusion free protocol for rational secret sharing,'' {\em IACR
  Eprint archive}, 2010.
\newblock \url{https://eprint.iacr.org/2010/250}.

\bibitem{araujo}
F.~Araujo, J.~Farinha, P.~Domingues, G.~C. Silaghi, and D.~Kondo, ``A maximum
  independent set approach for collusion detection in voting pools,'' {\em
  Journal of Parallel and Distributed Computing}, 2011.
\newblock \url{https://doi.org/10.1016/j.jpdc.2011.06.004}.

\bibitem{godel}
K.~G{\"o}del, {\em On Formally Undecidable Propositions of Principia
  Mathematica and Related Systems}.
\newblock Dover Publications, 1931.

\bibitem{church}
A.~Church, ``An unsolvable problem of elementary number theory,'' {\em American
  Journal of Mathematics}, vol.~58, no.~2, pp.~345--363, 1936.
\newblock \url{http://www.jstor.org/stable/2371045}.

\bibitem{turing}
A.~M. Turing, ``{On Computable Numbers With an Application to the
  Entscheidungsproblem},'' in {\em Proceedings of the London Mathematical
  Society}, 1937.

\bibitem{rosser}
B.~Rosser, ``Extensions of some theorems of gödel and church,'' {\em The
  Journal of Symbolic Logic}, vol.~1, no.~3, pp.~87--91, 1936.

\bibitem{rice}
H.~G. Rice, ``Classes of recursively enumerable sets and their decision
  problems,'' {\em Transactions of the American Mathematical Society}, vol.~74,
  no.~2, pp.~358--366, 1953.
\newblock \url{http://www.jstor.org/stable/1990888}.

\bibitem{kleene}
S.~C. Kleene, {\em Mathematical Logic}.
\newblock Dover Publications, 1967.

\bibitem{conway}
J.~H. Conway, ``On unsettleable arithmetical problems,'' {\em The American
  Mathematical Monthly}, vol.~120, no.~3, pp.~192--198, 2013.
\newblock
  \url{http://www.jstor.org/stable/10.4169/amer.math.monthly.120.03.192}.

\bibitem{hofstadter}
D.~R. Hofstadter, {\em Godel, Escher, Bach: An Eternal Golden Braid}.
\newblock New York, NY, USA: Basic Books, Inc., 1979.

\bibitem{filecoin}
J.~Benet, N.~Greco, D.~Dalrymple, M.~Zumwalt, E.~Miyazono, and {other
  contributors}, ``Filecoin: A decentralized storage network,'' 2014-2017.
\newblock \url{http://filecoin.io/filecoin.pdf}.

\bibitem{micali}
J.~Chen and S.~Micali, ``Algorand: The efficient and democratic ledger,'' 07
  2016.
\newblock \url{https://arxiv.org/pdf/1607.01341.pdf}.

\bibitem{Bentov:2014:PAE:2695533.2695545}
I.~Bentov, C.~Lee, A.~Mizrahi, and M.~Rosenfeld, ``Proof of activity: Extending
  bitcoin's proof of work via proof of stake [extended abstract]y,'' {\em
  SIGMETRICS Perform. Eval. Rev.}, vol.~42, pp.~34--37, Dec. 2014.
\newblock \url{https://eprint.iacr.org/2014/452.pdf}.

\bibitem{daian:2016:919}
P.~Daian, R.~Pass, and E.~Shi, ``Snow white: Provably secure proofs of stake.''
  Cryptology ePrint Archive, Report 2016/919, 2016.
\newblock \url{https://eprint.iacr.org/2016/919.pdf}.

\bibitem{bonneau}
J.~Bonneau, J.~Clark, and S.~Goldfeder, ``On bitcoin as a public randomness
  source,'' {\em IACR Cryptology ePrint Archive}, vol.~2015, p.~1015, 2015.
\newblock \url{https://eprint.iacr.org/2015/1015.pdf}.

\bibitem{satoshi:fees}
S.~Nakamoto, ``Email from april 2009 to mike hearn,'' 2009.
\newblock
  \url{https://bitcointalk.org/index.php?topic=149668.msg1596879#msg1596879}.

\bibitem{bitcoin:script}
{Bitcoin Wiki contributors}, ``Bitcoin wiki - script.''
\newblock \url{https://en.bitcoin.it/w/index.php?title=Script}.

\bibitem{ethereum}
G.~Wood, ``Ethereum: a secure, decentralised, generalised transaction ledger,''
  2014.
\newblock \url{http://gavwood.com/paper.pdf}.

\bibitem{michelson}
L.~M.~G. (Pseudonym), ``Michelson: the language of smart contracts in tezos,''
  2017.
\newblock \url{https://www.tezos.com/static/papers/language.pdf}.

\bibitem{simplicity}
R.~O'Connor, ``Simplicity: A new language for blockchains,'' 2017.
\newblock \url{https://blockstream.com/simplicity.pdf}.

\bibitem{solidity}
{Ethereum}, ``Solidity.''
\newblock
  \url{https://media.readthedocs.org/pdf/solidity/develop/solidity.pdf}.

\bibitem{ivy}
D.~Robinson, O.~Andreev, and T.~Arciery, ``Ivy: A declarative predicate
  language for smart contracts,'' 2017.
\newblock \url{https://chain.com/docs/1.2/ivy-playground/docs}.

\bibitem{schnorr}
C.~P. Schnorr, ``Efficient signature generation by smart cards,'' 1991.
\newblock
  \url{http://www.mi.informatik.uni-frankfurt.de/research/papers/schnorr.smartcardsig.1991.ps}.

\bibitem{robin:mast}
J.~Rubin, M.~Naik, and N.~Subramanian, ``Merkelized abstract syntax trees,''
  2014.
\newblock \url{http://www.mit.edu/~jlrubin/public/pdfs/858report.pdf}.

\bibitem{lau:mast}
J.~Lau, ``Bip 0114: Merkelized abstract syntax tree,'' 2016.
\newblock \url{https://github.com/bitcoin/bips/blob/master/bip-0114.mediawiki}.

\bibitem{lau:mastopcodes}
J.~Lau, ``Scripting system in merkelized abstract syntax tree,'' 2016.
\newblock
  \url{https://github.com/jl2012/bips/blob/mastopcodes/bip-mastopcodes.mediawiki}.

\bibitem{friedenbach:fastmerkle}
M.~Friedenbach, K.~Alm, and BtcDrak, ``Bip 98: Fast merkle trees,'' 2017.
\newblock \url{https://gist.github.com/maaku/41b0054de0731321d23e9da90ba4ee0a}.

\bibitem{friedenbach:merklebranchverify}
M.~Friedenbach, ``Bip 116: Merklebranchverify (consensus layer),'' 2017.
\newblock \url{https://gist.github.com/maaku/bcf63a208880bbf8135e453994c0e431}.

\bibitem{friedenbach:tailcall}
M.~Friedenbach, ``Bip 117: Tail call execution semantics (consensus layer),''
  2017.
\newblock \url{https://gist.github.com/maaku/f7b2e710c53f601279549aa74eeb5368}.

\bibitem{moser:covenants}
M.~Möser, I.~Eyal, and E.~G. Sirer, ``Bitcoin covenants,'' 2016.
\newblock \url{https://fc16.ifca.ai/bitcoin/papers/MES16.pdf}.

\bibitem{oconnor:covenants}
R.~O'Connor and M.~Piekarska, ``Enhancing bitcoin transactions with
  covenants,'' 2017.
\newblock \url{https://fc17.ifca.ai/bitcoin/papers/bitcoin17-final28.pdf}.

\bibitem{urbit}
C.~Yarvin, P.~Monk, A.~Dyudin, and R.~Pasco, ``Urbit: A solid-state
  interpreter,'' 2016.
\newblock \url{http://media.urbit.org/whitepaper.pdf}.

\bibitem{eventsource}
Mozilla and individual contributors, ``Server-sent events,'' 2011.
\newblock
  \url{https://developer.mozilla.org/en-US/docs/Web/API/Server-sent_events}.

\bibitem{wiki:verifiability}
{Wikipedia contributors}, ``{Verifiability, not truth} --- wikipedia{,} the
  free encyclopedia.''
\newblock
  \url{https://en.wikipedia.org/wiki/Wikipedia:Verifiability,_not_truth}.

\bibitem{huberman}
B.~A. Huberman and R.~M. Lukose, ``Social dilemmas and internet congestion,''
  {\em Science}, vol.~277, no.~5325, pp.~535--537, 1997.
\newblock
  \url{https://www.researchgate.net/publication/317903463_Social_Dilemmas_and_Internet_Congestions}.

\bibitem{hess}
C.~Hess and E.~Ostrom, {\em Understanding Knowledge as a Commons: From Theory
  to Practice}.
\newblock The MIT Press, 2007.

\bibitem{hardin}
G.~Hardin, ``The tragedy of the commons,'' {\em Science}, vol.~162, no.~3859,
  pp.~1243--1248, 1968.

\bibitem{lloyd}
W.~F. Lloyd, {\em Two lectures on the checks to population}.
\newblock The University of Oxford, 1833.

\bibitem{fuster}
M.~Fuster, {\em Governance of online creation communities: Provision of
  platforms for participation for the building of digital commons}.
\newblock PhD thesis, European University Institute, 2009.
\newblock \url{https://goo.gl/aH7e6B}.

\bibitem{benkler}
Y.~Benkler, ``Coase's penguin, or, linux and \textit{The Nature of the Firm},''
  {\em The Yale Law Journal}, vol.~112, pp.~369--446, 2002.
\newblock
  \url{http://www.yalelawjournal.org/article/coases-penguin-or-linux-and-the-nature-of-the-firm}.

\bibitem{lockss:about}
L.~O. C. K. S.~S. program, ``What is lockss?.''
\newblock \url{https://www.lockss.org/about/what-is-lockss/}.

\bibitem{lockss:principles}
L.~O. C. K. S.~S. program, ``Preservation principles.''
\newblock \url{https://www.lockss.org/about/principles/}.

\bibitem{ipfs}
J.~Benet, ``{IPFS - Content Addressed, Versioned, P2P File System (DRAFT 3)},''
  2015.
\newblock
  \url{https://ipfs.io/ipfs/QmR7GSQM93Cx5eAg6a6yRzNde1FQv7uL6X1o4k7zrJa3LX/ipfs.draft3.pdf}.

\bibitem{sia}
D.~Vorick and L.~Champine, ``Sia: Simple decentralized storage,'' 2014.
\newblock \url{https://www.sia.tech/whitepaper.pdf}.

\bibitem{storj}
S.~Wilkinson, T.~Boshevski, J.~Brandoff, J.~Prestwich, G.~Hall, P.~Gerbes,
  P.~Hutchins, and C.~Pollard, ``Storj: A peer-to-peer cloud storage network,''
  2016.
\newblock \url{https://storj.io/storj.pdf}.

\bibitem{swarm1}
V.~Trón, A.~Fischer, D.~A. Nagy, Z.~Felföldi, and N.~Johnson, ``Swap, swear
  and swindle: Incentive system for swarm,'' 2016.
\newblock
  \url{http://swarm-gateways.net/bzz:/theswarm.eth/ethersphere/orange-papers/1/sw^3.pdf}.

\bibitem{swarm2}
V.~Trón, A.~Fischer, and N.~Johnson, ``Smash-proof: auditable storage for
  swarm secured by masked audit secret hash,'' 2016.
\newblock
  \url{http://swarm-gateways.net/bzz:/theswarm.eth/ethersphere/orange-papers/2/smash.pdf}.

\end{thebibliography}
	
\end{document}